\documentclass[journal,hideappendix]{vgtc} 

\onlineid{1306}

\vgtccategory{Research}

\title{\textbf{VACP}: Visual Analytics Context Protocol}

\author{%
    \authororcid{Tobias Stähle}{0009-0001-5983-8807},
    \authororcid{Péter Ferenc Gyarmati}{0009-0006-6122-2709},
    \authororcid{Thilo Spinner}{0000-0002-1168-1804},\\
    \authororcid{Rita Sevastjanova}{0000-0002-2629-9579},
    \authororcid{Dominik Moritz}{0000-0002-3110-1053},
    \authororcid{Mennatallah El-Assady}{0000-0001-8526-2613}
}

\authorfooter{
    \item
    Tobias Stähle, Péter Ferenc Gyarmati, Thilo Spinner, Rita Sevastjanova\\and Mennatallah El-Assady are with ETH Zürich.
    \item
    Dominik Moritz is with Carnegie Mellon University. 
}

\usepackage{xcolor}
\usepackage[normalem]{ulem}
\usepackage{etoolbox} 
\usepackage{tabularx}

\newbool{showrevisions}
\setbool{showrevisions}{false} 

\definecolor{addcolor}{RGB}{200,0,0}    
\definecolor{delcolor}{RGB}{200,0,0}    

\ifbool{showrevisions}{
  \newcommand{\add}[1]{\textcolor{addcolor}{#1}}
  \newcommand{\remove}[1]{\textcolor{delcolor}{\sout{#1}}}
  \newcommand{\replace}[2]{\textcolor{delcolor}{\sout{#1}} \textcolor{addcolor}{#2}}
  \newcommand{\showRevisionUpdate}[1]{\textcolor{addcolor}{#1}}
}{
  \newcommand{\add}[1]{#1}
  \newcommand{\remove}[1]{}
  \newcommand{\replace}[2]{#2}
  \newcommand{\showRevisionUpdate}[1]{}
}

\newenvironment{addblock}{%
  \ifbool{showrevisions}{%
    \color{addcolor}
  }{%
  }%
}{%
  \ifbool{showrevisions}{}{\endcomment}%
}

\newcommand{\iviaparagraph}[1]{\refstepcounter{paragraph}\noindent\textbf{#1\ ---}\label{par:\theparagraph}}

\usepackage{xspace}

\usepackage{inlinegraphicx}

\usepackage{overpic}

\usepackage{wrapfig2}

\usepackage{fontawesome5}

\usepackage{tcolorbox}
\usepackage{tikz}

\usepackage{colortbl}
\usepackage{makecell}

\makeatletter
\newtcbox{\kwColorBox}[1][]{on line,fontupper=\footnotesize\sffamily\bfseries\small,boxrule=0.5pt,arc=2pt,coltext=#1,colback=#1!10!white,colframe=#1,boxsep=0pt,left=1.5pt,right=1.5pt,top=1.5pt,bottom=1.5pt}

\newtcbox{\kwColorBoxSpecial}[1][]{on line,fontupper=\footnotesize\sffamily\bfseries\small,boxrule=0.5pt,arc=2pt,coltext=white,colback=#1,colframe=#1,boxsep=0pt,left=1.5pt,right=1.5pt,top=1.5pt,bottom=1.5pt}

\newtcbox{\kwColorBoxSpecialContext}[1][]{on line,fontupper=\footnotesize\sffamily\bfseries\small,boxrule=0.5pt,arc=2pt,coltext=white,colback=#1,colframe=#1,boxsep=0pt,left=1.5pt,right=1.5pt,top=1.5pt,bottom=1.5pt}
\makeatother

\newcommand{\kw}[2]{%
    \begin{kwColorBox}[#2]%
    {#1}%
    \end{kwColorBox}%
}
\newcommand{\kwSpecial}[2]{%
    \begin{kwColorBoxSpecial}[#2]%
    {#1}%
    \end{kwColorBoxSpecial}%
}
\newcommand{\kwSpecialContext}[2]{%
    \begin{kwColorBoxSpecialContext}[#2]%
    {#1}%
    \end{kwColorBoxSpecialContext}%
}

\definecolor{ColorUser}{HTML}{A42C2C}
\definecolor{ColorChallenge}{HTML}{A02FA5}
\definecolor{ColorLmChallenge}{HTML}{6B4E90}
\definecolor{ColorTask}{HTML}{408E2F}
\definecolor{ColorLmTask}{HTML}{AA7A39}
\definecolor{ColorWidget}{HTML}{4F85C3}
\definecolor{ColorPrinciple}{HTML}{99BDB2}
\definecolor{ColorLayer}{HTML}{8F8F8F}
\definecolor{ColorPrincipleDynamics}{HTML}{61750C}

\newcommand{\rawlmprincipledonotuse}[1]{\kw{#1}{ColorPrinciple}}
\newcommand{\deflmprinciple}[1]{\rawlmprincipledonotuse{\phantomsection\label{lmprinciple:#1}#1}}
\newcommand{\reflmprinciple}[1]{\hyperref[lmprinciple:#1]{\rawlmprincipledonotuse{#1}}}

\newcommand{\rawlayerdonotuse}[1]{\kw{#1}{ColorLayer}}
\newcommand{\deflayer}[1]{\rawlayerdonotuse{\phantomsection\label{lmlayer:#1}#1}}
\newcommand{\reflayer}[1]{\hyperref[lmlayer:#1]{\rawlayerdonotuse{#1}}}

\newcommand{\rawcontextdonotuse}[1]{\kwSpecialContext{#1}{ColorPrinciple}}
\newcommand{\refcontext}[1]{\hyperref[lmprinciple:#1]{\rawcontextdonotuse{#1}}}
\newcommand{\rawdynamicdonotuse}[1]{\kwSpecial{#1}{ColorPrincipleDynamics}}
\newcommand{\refdynamic}[1]{\hyperref[lmprinciple:#1]{\rawdynamicdonotuse{#1}}}

\newcommand{\rawuserdonotuse}[1]{\kw{#1}{ColorUser}}

\newcommand{\refuser}[1]{\hyperref[user:#1]{\rawuserdonotuse{#1}}}

\usepackage{tikz}
\usepackage{tcolorbox}
\usepackage{graphicx} 
\usepackage{subcaption}

\definecolor{VACP_color}{HTML}{C6D38F}
\definecolor{myGlobal}{HTML}{1b9e77}
\definecolor{myAxis}{HTML}{7570b3}

\newcommand{\DrawLine}{%
    \begin{tikzpicture}
    \path[use as bounding box] (0,0) -- (\linewidth,0);
    \draw[color=red!75!black,dashed,dash phase=2pt]
    (0-\kvtcb@leftlower-\kvtcb@boxsep,0)--
    (\linewidth+\kvtcb@rightlower+\kvtcb@boxsep,0);
    \end{tikzpicture}%
}

\newcommand{\dimension}[5]{
    \vspace{-0.25em}
    \begin{tcolorbox}[
        fonttitle=\bfseries,
        coltitle=black,
        colbacktitle=#2!40,
        colback=#2!20, 
        colframe=#2,
        title={#1 \hfill #4}, 
        after skip=0.35em,
        left=3pt, right=3pt, top=1pt, bottom=1pt,
        boxsep=0pt,
        middle=0.4em,
        toptitle=1pt, bottomtitle=1pt,
        sharp corners=all,
        boxrule=0mm, leftrule=1mm
    ]

    #3

    \end{tcolorbox}%
    \noindent#5
}

\usepackage{soul}

\newcommand{\underlineColor}[2]{%
  \ifbool{showrevisions}{%
    \textcolor{#1}{#2}%
  }{%
    \expandafter\underlineColorClean\expandafter{#2}{#1}%
  }%
}

\newcommand{\underlineColorClean}[2]{%
  {\markoverwith{\textcolor{#2}{\rule[-0.5ex]{2pt}{0.4pt}}}\uline{#1}}%
}

\definecolor{VACPColor}{HTML}{5E8A7D}

\newcommand{\rawvacpRequirement}[3]{%
\underlineColor{VACPColor}{#3} ({\leavevmode\color{VACPColor}\emph{ch#1}: \leavevmode\color{VACPColor}{\textbf{#2}}})%
}

\newcommand{\rawvacpRequirementBegin}[1]{%
\underlineColor{VACPColor}{#1}
}
\newcommand{\rawrefvacpRequirement}[1]{%
{\leavevmode\color{VACPColor}\emph{ch#1}}%
}
\newcommand{\defrequirement}[3]{\rawvacpRequirement{\phantomsection\label{vacpreq:#1}#1}{#2}{#3}}

\newcommand{\defrequirementBegin}[1]{#1}
\newcommand{\refrequirement}[1]{\hyperref[vacpreq:#1]{\rawrefvacpRequirement{#1}}}%
\newcommand{\refrequirementDetails}[2]{\rawvacpRequirementBegin{#2}(\hyperref[vacpreq:#1]{\rawrefvacpRequirement{#1}})}%

\abstract{%
    The rise of AI agents introduces a fundamental shift in Visual Analytics (VA), in which agents act as a new user group. 
    Current agentic approaches---based on computer vision and raw DOM access---fail to perform VA tasks accurately and efficiently.
    This paper introduces the Visual Analytics Context Protocol (VACP), a framework designed to make VA applications ``\textit{agent-ready}'' that extends generic protocols by explicitly exposing application state, available interactions, and mechanisms for direct execution.
    To support our context protocol, we contribute a formal specification of AI agent requirements and knowledge representations in VA interfaces. We instantiate VACP as a library compatible with major visualization grammars and web frameworks, enabling augmentation of existing systems and the development of new ones.
    Our evaluation across \replace{representative}{a compelling set of} VA tasks demonstrates that VACP-enabled agents achieve higher success rates in interface interpretation and execution compared to current agentic approaches, while reducing token consumption and \replace{latency}{completion time}.
    VACP closes the gap between human-centric VA interfaces and machine perceivability, ensuring agents can reliably act as collaborative users in VA systems. More information under \href{https://vacp.dev/}{https://vacp.dev}. %
}


\teaser{
    \centering
    \includegraphics[width=\linewidth, alt={A view of a city with buildings peeking out of the clouds.}]{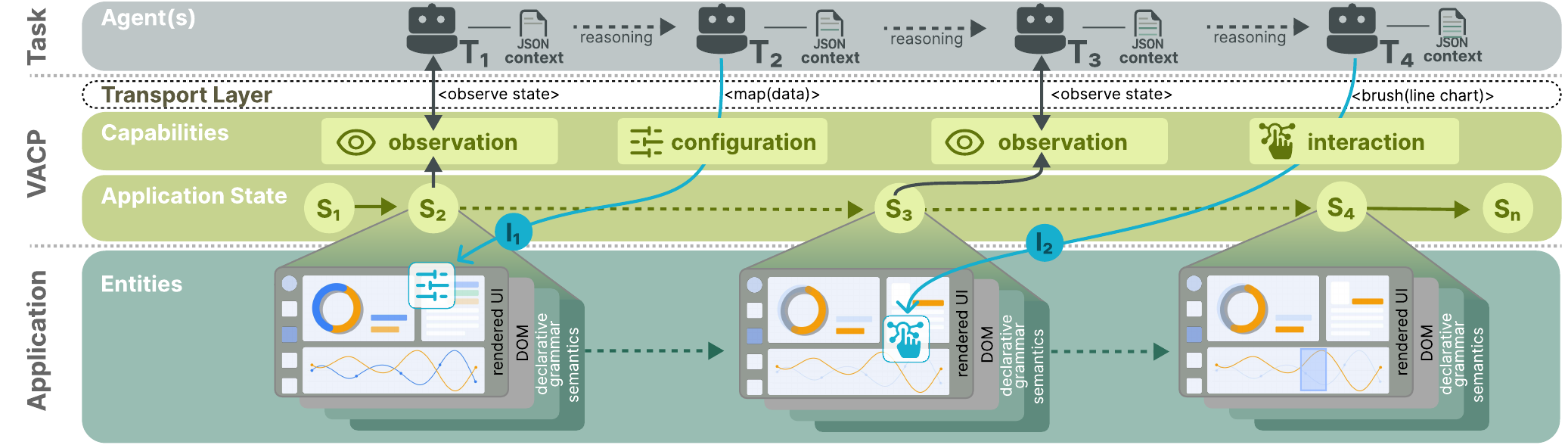}
    \caption{%
       AI agents interacting with Visual Analytics (VA) interfaces using \textbf{VACP} (Visual Analytics Context Protocol). VACP maintains the \textbf{application state} and available \textbf{interactions} informed through the different types of knowledge representation of the \textbf{VA entities}. To perform analytical tasks in the interface, an \textbf{AI agent} can use the provided capabilities to perceive the VA interface and understand possible interactions by using VACP. Further, it can interact with the VA interface through the provided interactions. Interactions (e.g., $I_1$ or $I_2$) manipulate the interface appearance,  triggering a state transition (e.g., $S_2 \rightarrow S_3$) in the application within the VACP layer. %
    }
    \label{fig:teaser}
}

\graphicspath{{figs/}{figures/}{pictures/}{images/}{./}} 

\usepackage{tabu}                      
\usepackage{booktabs}                  
\usepackage{lipsum}                    
\usepackage{mwe}                       
\usepackage{ccicons}                   

\usepackage{mathptmx}                  

\begin{document}
\crefformat{footnote}{#2\footnotemark[#1]#3}

\firstsection{Introduction}

\maketitle
Visual Analytics (VA) systems are traditionally designed with the implicit assumption that \textit{\textbf{human users}} perceive visually encoded data and cognitively process it to observe patterns. 
To enhance analytical workflows, mixed-initiative VA systems use machine learning models and agents to automate subtasks or provide guidance~\cite{ceneda2017characterizingGuidance, sperrle_2021_learning}.  These agents mainly serve as an additional component of VA systems.

Unlike traditional integrated models, recent advances in agentic AI unlock a new paradigm: \textit{\textbf{AI users}} --- autonomous agents directly operating interactive web interfaces~\cite{wang2024autonomousAgents}, e.g., autonomously performing selection, filtering, and drill-down operations~\cite{battle2019characterizingVA} to answer analytical questions~\cite{kartha2025dashboardQA}. In such contexts, the success of AI agents is contingent on their ability to operate and interact directly with the visual interface. If realized, VA interfaces supporting human and AI agents side by side can unlock powerful potential, enabling real-time collaboration in shared, dynamic analytical environments.

Despite this potential, current agentic approaches, such as AI agents powered by recent multimodal foundation models like Gemini~\cite{gemini2023geminiFamily}, GPT-5.2~\cite{openAI2025GPT52}, and Claude Sonnet 4.5~\cite{Anthropic2025ClaudeSonnet45}, attempt to navigate interfaces visually using Computer Vision or raw HTML parsing~\cite{Antropic_ComputerUse_2024, OpenAI_ComputerUseAgent_2025, BrowserUse_2025, wang2024autonomousAgents}.
By treating agents as human simulators, these approaches \textit{struggle to interact with VA applications}. They suffer from poor information retention~\cite{kartha2025dashboardQA} and frequently fail to execute interactions correctly.
In early evaluations testing agents on real-world analytical tasks (\cref{sec:evaluation}), we observed low task completion rates driven by varied failure modes, imprecise data selection, and significant interaction execution errors. 

We identify this difficulty as a critical ``\textbf{\textit{interface mismatch}}'' that currently impedes progress. Existing VA systems are optimized for human perception (rendered pixels) and human action (mouse events), creating an environment that is hostile to AI agents. We argue that relying on pixel-level interpretation for logical tasks is \remove{fundamentally }flawed. It introduces unnecessary latency, stochasticity, and hallucination, forcing a high-intelligence model to operate through a low-fidelity ``\textit{keyhole}.''

Breaking free from this bottleneck unlocks new possibilities for operating mixed-initiative systems.
By bypassing visual approximations and exposing underlying logic and data states, we establish a direct connection between VA system and AI agent, allowing seamless retrofitting of legacy applications to support agentic accessibility tools and enabling human and AI agents to collaborate within a shared analytical context. Furthermore, the machine-readable interface serves as basis for rich programmatic evaluation environments, where researchers can systematically simulate, test, and benchmark interactive workflows.
%
%

Drawing on existing VA literature and recent work on intelligent agents in VA~\cite{staehle2025agentDesignSpace, monadjemi_agentBased_2023}, we identify that agents require access to information about the application's current state, available interaction options, and the ability to execute interactions. Further, we investigate which interface layers encode this knowledge to model application state and interaction representation in a way that balances semantic richness with token efficiency for the AI agent while considering limitations of current AI agents such as context window size~\cite{hong2025context, zhang2025RecursiveLM}.

%

Recent general-purpose standards, such as the Model Context Protocol (MCP)~\cite{Antropic_2025_MCP}, expose tools to AI agents but lack the domain-specific semantic grounding needed for complex, stateful VA. 
This paper introduces the \textbf{Visual Analytics Context Protocol} (VACP), which makes current and future VA systems ``\textit{AI-agent-ready}.'' VACP uses MCP as the communication substrate, and operates above its transport layers, augmenting agent context by exposing the hierarchical data structures and precise interaction spaces unique to VA systems.
By exposing a structured application state and clear execution endpoints, VACP transforms VA interfaces into an environment where AI agents can perform analytical tasks reliably and accurately. This allows developers to retrofit existing systems and build new agent-ready VA systems. To our knowledge, \textit{VACP is the first context protocol to establish a direct interaction bridge between the VA interface and AI agents as a new synthetic user type.} 
Specifically, we contribute: 

\noindent\textbf{(1) Specification of Challenges for AI Agents' Accessibility in VA:} A specification of the knowledge representation non-human actors require to navigate VA systems, combining established interaction taxonomies~\cite{yi2007interaction, dimara2020interaction, brehmer2013typology, heer2012interactive} with the different levels of knowledge representation in VA applications~\cite{sperrle2021coadaptive}.
\\\noindent\textbf{(2) Visual Analytics Context Protocol:} A definition of principles for context protocols in interactive, data-rich environments. Based on these, we formally describe VACP, which exposes a context-rich application state, available interactions, and mechanisms for direct execution.
\remove{\\}\noindent\remove{\textbf{The VACP Library:} We provide an instantiation of VACP as a TypeScript library that lets developers build agent-ready VA applications at design time with support for modern web frameworks.
The architecture supports vanilla JavaScript and frameworks such as D3.js, with built-in default compatibility for higher-level declarative formats (Vega~\mbox{\cite{satyanaeayan2016vega}}, Vega-Lite~\mbox{\cite{satyanarayan2017vegaLite}}, and Mosaic's vgplot~\mbox{\cite{heer2024mosaic}}). Its flexible design lets developers extend VACP-provided information to enrich the agent's context.}
\remove{\\}\noindent\remove{\textbf{A Comparative Evaluation of VACP-enabled Agents:} We compare VACP-enabled agents with state-of-the-art methods and find that VACP agents achieve higher task completion rates while significantly reducing token consumption and execution time. 
Additionally, when existing vision-based approaches are augmented with VACP, agents can reliably perform analytical tasks and independently verify visual outputs using their vision capabilities.}
\\\noindent\add{\textbf{(3) VACP Library and Evaluation:} An instantiation of VACP as a TypeScript library, supporting modern web frameworks and ensuring built-in default compatibility for higher-level declarative formats.
Our evaluation compares VACP-enabled agents with state-of-the-art methods and shows that VACP agents achieve higher task completion rates while significantly reducing token consumption and execution time. }
%
%

By replacing ad-hoc, heuristic observation with robust, formal communication, VACP lays the foundation for turning AI agents from passive observers into reliable analytical collaborators. Released as open-source software\footnote{\label{foot:github-link}Available as open-source at \faGithub:~\href{https://github.com/ETH-IVIA-Lab/VACP}{github.com/ETH-IVIA-Lab/VACP}}, \replace{VACP empowers VA developers}{VA developers can use VACP} to \textit{build new or supercharge existing VA systems with interfaces designed from the ground up for AI agent accessibility}.

\section{Background \& Related Work}
\iviaparagraph{Web Agents and Browser Automation}
As LLMs evolved from passive text processors into active agents, research on autonomous web navigation accelerated~\cite{wang2024autonomousAgents}.
Early automation relied on rigid scripts such as Selenium~\cite{selenium_project} or on reinforcement learning in simplified environments like \textit{World of Bits}~\cite{shi2017worldOfBits}.
Instead, recent agentic approaches exploit LLMs' semantic reasoning over the Document Object Model (DOM)~\cite{whatwh2026DOM} or screenshots to execute tasks dynamically~\cite{gur2024realWorldWebAgent}.
Chrome's recently proposed WebMCP~\cite{webMCP2026} embeds a JavaScript interface that exposes application functionality as tools for AI browser agents to invoke.
However, these approaches suffer from a critical \textit{Interface Mismatch} in VA.
DOM-based agents struggle because modern VA applications rely on \texttt{<canvas>} or complex SVG that exceeds LLM context windows~\cite{gur2024realWorldWebAgent}.
To a standard DOM parser, a D3.js scatterplot is a \textit{semantic black box}: thousands of anonymous path elements or a single opaque pixel buffer, lacking the descriptors needed for reasoning.
Overcoming the context-window bottleneck would require aggressive DOM pruning, which strips the interface logic needed to understand it properly.
Vision-based agents (e.g., \textit{pixels-to-actions}~\cite{shaw2023PixelsToUI} and computer-use agents~\cite{Antropic_ComputerUse_2024, OpenAI_ComputerUseAgent_2025}) better mimic human perception but lack the coordinate precision needed in dense data environments.
As Kartha et al.~\cite{kartha2025dashboardQA} note, this yields high hallucination rates with agents misinterpreting visual overlaps or failing to distinguish subtle color encodings. \add{Work on visualization literacy evaluates how well multimodal LLMs interpret visualizations to assist users \cite{choe_enhancingLiteracy_2025}, including under varying color, shape, and text~\cite{valentim_PlotThickens_2025}.
These studies find that current LLMs handle overall chart comprehension~\cite{lo_howGood_2025} and detect chart hierarchies and data trends~\cite{pandey_benchmarkingVLM_2025} well, yet struggle with data retrieval and anomaly detection in dense visualizations~\cite{bendeck_empirical_2025,pandey_benchmarkingVLM_2025}. }%
While benchmarks \add{on visualization literacy and general web tasks} like WebArena~\cite{zhou2024WebArena} and Mind2Web~\cite{deng2023mind2web} exist\remove{ to assess the performance of AI agents in general web tasks}, to our knowledge none target the interactive VA domain.

\vspace{.25em}
\iviaparagraph{Visualization Accessibility for Agent Perception}
We posit that exposing VA systems to AI agents parallels making visualizations accessible to vision-impaired users.
In both, consumers cannot rely on visual gestalt and need a structured, traversable alternative~\cite{elavsky2023data,jones_customization_2024}, while AI agents additionally require explicit semantic representations. 
\\
Visualization accessibility has matured from simple alt-text generation to rich semantic structures.
Zong et al.~\cite{zong2022richScreenReaders} and Lundgard and Satyanarayan~\cite{lundgard2022accessibleVis} argue that true accessibility requires exposing not only statistical facts but also the hierarchical and semantic relationships among data, encodings, and annotations.
\\
In VA, Sacha et al.~\cite{sacha2016uncertainty} note that resolution and clutter create uncertainty for users --- uncertainty that, for agents, is exacerbated by the stochastic nature of vision-language models.
Declarative grammars like \add{Vega~\cite{satyanaeayan2016vega}, }Vega-Lite~\cite{satyanarayan2017vegaLite} and Mosaic's vgplot~\cite{heer2024mosaic,heer_mosaic_2026} provide a baseline here, as they explicitly map data to visual channels.
Das et al.~\cite{das2025chartsOfThought} confirm that multimodal LLMs interpret visualizations better when given such structured data. \add{Other approaches pair LLMs with the underlying data and visualization specifications to enable natural language interfaces~\cite{jobst_EvaluatingNLI_2026}, such as VizAbility~\cite{gorniak_Vizability_2024}.}
\\
A key distinction remains: existing accessibility layers are designed for \textit{consumption} --- reading the visualizations --- and rarely expose the interface \textit{affordances} for \textit{operation} (e.g., that brushing interval $A$ filters view $B$).
Kim et al.~\cite{kim2021accessibleVisDesignSpace} stress the need for non-visual interaction modalities as interactive visualizations grow more complex, and the absence of this meta-information leads to what Gillmann et al.~\cite{gillmann2023uncertainty} call usage uncertainty.
By extending VA's read-only semantic schemas into a read-write protocol, VACP lets AI agents actively manipulate system state, treating the agent as a user with specific ``\textit{accessibility}'' needs.

\vspace{.25em}
\iviaparagraph{Interactions in Visualizations}
The importance of interaction for effective VA applications is widely recognized~\cite{yi2007interaction, pike2009scienceInteraction} \add{, and interaction provenance has been analyzed to study user behavior~\cite{north_AnaProvenance_2011}}.
Existing taxonomies categorize interactions by user intent~\cite{yi2007interaction} and interactive dynamics~\cite{heer2012interactive}, providing the vocabulary to describe what an agent intends to do.
While interaction-design guidelines exist~\cite{sedig2013interactionDesign, bach2023dashboard}, implementation is often tightly coupled to the rendering engine~\cite{zhao2025libra}. %
\\
Dimara and Perin~\cite{dimara2020interaction} argue for distinguishing an interaction's \textbf{intent} from its \textbf{physical mechanism}, decoupling utility from implementation, and enabling evaluation of techniques independent of input modalities.
For mixed-initiative VA~\cite{ceneda2017characterizingGuidance}, where agents proactively support users, this requires a formal definition of the interaction space.
Declarative grammars like Vega-Lite model the \textit{intent} of interactions internally (via Signals and Selections), but many lack an \textbf{Interaction Gateway}, i.e., a semantic Application Programming Interface (API) that lets external agents discover and trigger intents directly without passing through the UI layer.
Since the physical mechanism (e.g., simulating mouse drags) is a hindrance for current AI agents~\cite{zhou2024WebArena}, recent work on agentic tool use suggests they are more effective at the intent level (API/tool calls)~\cite{schick2023toolformer}.
Advancements such as the Model Context Protocol (MCP)~\cite{Antropic_2025_MCP} lower this bound by exposing predefined tools in a structured format and tools like Embedding Atlas~\cite{ren2025embeddingAtlas} already use MCP. Yet these approaches could benefit from a more efficient protocol.
\\
We conceptualize agent-VA interactions in the UI as two stages.
Stage~1 covers operating predefined interfaces (e.g., selecting, filtering, highlighting) to generate insights within fixed constraints, whereas Stage~2 involves authoring or reconfiguring interactive dashboards.
Although dashboard generation and composition are popular research topics~\cite{dibia2023lida}, an agent's ability to (re)configure a dashboard has limited value without subsequent interaction to refine findings.
This paper focuses on Stage~1 interactions as the operational foundation for future complex authoring tasks, with a Stage~2 use case provided in the supplemental material.

\section{Knowledge and Application State Access in VA}
\label{sec:KnowledgeApplicationStateAccess}
Building a gateway that lets AI agents perceive and operate within interactive VA applications requires identifying their functional components.
Drawing on HCI and Information Visualization research~\cite{bach2023dashboard,brehmer2013typology,dimara2020interaction,heer2012interactive,zhao2025libra}, we decompose VA interfaces into two operational dimensions: \emph{Knowledge Representation} and \emph{Interactive Functionality}.
This decomposition defines the state space an AI agent must perceive to reason effectively about the analytics process, and we analyze it to identify challenges~(\refrequirement{XX}) for providing AI agents accessibility in VA interfaces.

\vspace{.25em}
\iviaparagraph{Knowledge Representation and Visual Composition}
The primary function of VA interfaces is to represent knowledge, serving as a visual proxy for the underlying data model.
At its core, a VA interface consists of at least one visualization view whose graphical elements encode data attributes as visual tokens (e.g., marks).
Modern VA tools often compose multiple views into complex interfaces~\cite{bach2023dashboard}.
For an AI agent, understanding such an interface takes more than detecting individual visualizations: it requires \defrequirement{01}{Understanding Interface Logic}{parsing the structural logic or ``grammar'' of how views are spatially arranged and logically linked to represent high-dimensional knowledge}. %

\vspace{.25em}
\iviaparagraph{Interactivity, Operational Dynamics, and Feedback}
Beyond static representation, the interface is defined by its Interactivity and Interaction Feedback.
Literature in VA describes interactions abstractly in terms of user intent rather than technical implementation in web interfaces~\cite{heer2012interactive, munzner2014analysisAndDesign, lu2017interaction}.
Following Munzner’s What-Why-How framework~\cite{munzner2014analysisAndDesign}, these \defrequirementBegin{interactive elements (lasso, zoom handles, or sliders) operationalize the user's or agent's intent} (the “Why”).
As Heer and Shneiderman~\cite{heer2012interactive} note, they drive transitions between data states (e.g., filtering) and visual states (e.g., encoding changes).
\add{When using VA interfaces,} AI agents must \defrequirement{02}{Affordance Identification}{identify and interact with these actuators to conduct or steer analysis}.

High-level analytic interactions consist of sequential low-level input events (e.g., mouse events), and executing operations such as brushing a specific data cluster demands highly accurate spatial coordination.
AI agents must therefore not only decide what action to take but also \defrequirement{03}{Accurate Interaction}{synthesize and execute these fine-grained, coordinate-based steps to reliably update the data state}.

Effective interaction requires perceptible responsiveness~\cite{pike2009scienceInteraction}.
To close the interaction loop, VA systems provide visual feedback (e.g., highlighted selections, tooltips) confirming how actions change the system state~\cite{lu2017interaction}.
This is critical for an AI agent: to interact reliably, it must \defrequirement{04}{Interaction Feedback}{verify that its actions have been successfully registered and processed by the application}.

\vspace{.25em}
\iviaparagraph{Knowledge Layers}Data-rich, interactive VA environments pose distinct challenges for autonomous AI agents beyond general web navigation.
VA application states are transient and multifaceted, and many interfaces rely on users' intuition and prior knowledge to discover interactions.
While some interfaces provide tutorials~\cite{bach2023dashboard}, many assume implicit expectations and user expertise.
Because LLM-based agents only know what appears in their training corpus, \defrequirement{05}{Providing Context}{\replace{we cannot assume they possess the necessary knowledge or implicit expectations about VA interactions}{they need to be informed with the necessary knowledge and possible implicit expectations about VA interactions}}.

The ``correct'' interpretation of knowledge and interaction possibilities depends strongly on the interplay among the underlying data distribution, the user's analytical task, and the provided information.

Prior work~\cite{martins2025talkingToData, wang2025DeclarativeLLMInterfaces} shows performance to scale with the \defrequirementBegin{structural clarity of the environment.} However, \defrequirementBegin{there is no standardized context protocol for an agent to query the ``holistic state''} of a VA system across knowledge representation layers to interact with the interface.
Moving from simple UI automation to collaborative analytics thus requires a \refrequirementDetails{01}{unified representation that synthesizes perceptual, structural, and semantic knowledge}.
We identify four interlinked VA knowledge representation layers (see \cref{fig:knowledgeLayers}): 
\deflayer{L1}~pixel-based, 
\deflayer{L2}~DOM-based, 
\deflayer{L3}~declarative grammar-based, and 
\deflayer{L4}~production logic-based.

\subsection{\protect\reflayer{L1}~Pixel-based Representation}

\begin{figure}[t]%
  \centering %
  \includegraphics[width=\columnwidth, alt={Overview of knowledge representation layers in VA applications, agents' accessibility, and VACP functions to perceive the VA interface on each level and interaction functionalities. Each layer encodes and holds an abstraction of the knowledge that informs about details such as the data to analyze, data encoding into visual variables, available interactions, and reasons for specific interface functionalities based on the production logic and design decisions. Checkboxes indicate the level of knowledge representation at which state-of-the-art AI web agents and AI agents using VACP can advance. While using VACP to interact with a VA application, it is not mandatory that the agent consumes all of the layers at once.}]{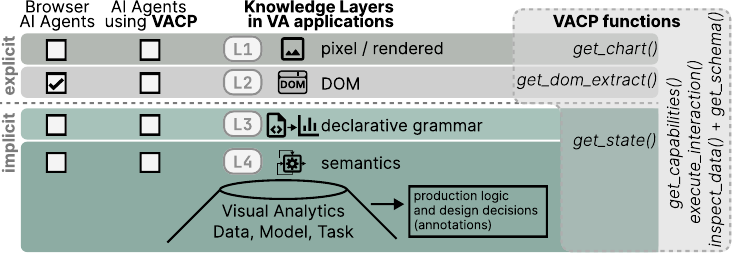}
  \vspace{-2em}
  \caption{%
  \showRevisionUpdate{[Figure updated] }Overview of knowledge representation layers in VA applications, agents' accessibility, and VACP functions for perceiving the VA interface at each level and supporting interactions. Each layer abstracts the knowledge, including the data to be analyzed, data encoding, and interface functionalities defined by production logic and design decisions. %
  }
  \vspace{-2em}
  \label{fig:knowledgeLayers}
\end{figure}

\noindent\textit{\textbf{Definition and Scope.}} In VA, the most practical knowledge representation is the rendered (usually 2D) interface that can be visually perceived.
In web-based VA applications, this is the rendered DOM, i.e., the final product.
Within the VA design process, it is the highest level of knowledge abstraction, summarizing all information visually. 
When data or visualization logic changes, this representation must be re-rendered.
\textit{\textbf{Format and Structure.}} The pixel-based representation is unstructured.
Unlike the underlying layers (\reflayer{L2}~--~\reflayer{L4}, see \cref{fig:knowledgeLayers}), it can only convey as much information as there are pixels on the screen.
\textit{\textbf{Encoded Knowledge and Added Value.}} Despite being unstructured, this representation can \defrequirementBegin{extract the application layout and the position of visual elements.} In data visualizations~\cite{bach2023dashboard}, it reveals visually \defrequirementBegin{detectable data patterns, such as clusters, outliers, trends, high-density regions, and differences among data points} via distinct visual encodings (e.g., color, shape).
\textit{\textbf{Interaction Capabilities.}} Many VA systems implement interactions for further exploration, but in the pixel-based representation, these are usually not explicitly described.
Unless the visual interface explicitly lists and describes them (e.g., in text or explicit affordances), users and agents must \refrequirementDetails{02}{infer available interactions from experience and implicit expectations}.
\textit{\textbf{Agent Accessibility.}} An AI agent can access this layer by \defrequirementBegin{taking a screenshot of the current application interface.} To interpret it, \defrequirement{06}{Modality Access}{the agent needs computer vision capabilities}.

\subsection{\protect\reflayer{L2}~DOM-based Representation}

\noindent\textit{\textbf{Definition and Scope.}} In web applications, the Document Object Model (DOM) is a browser-managed logical tree representing the application's structure, element roles, and content.
It abstracts VA knowledge from visual grammar and the developer’s production logic, capturing the interface's live state for rendering and functionality.
\textit{\textbf{Format and Structure.}} The DOM is typically expressed in the text-based, \textsc{XML}-structured format---HTML---that defines the application’s component hierarchy.
Tags can contain text and code.
The DOM derived from HTML is dynamic and can be modified with JavaScript.
\textit{\textbf{Encoded Knowledge and Added Value.}} The DOM layer adds value beyond the pixel-based representation.
It maintains the \reflayer{L1} knowledge representation but augments it with retrievable data values, exact numbers, and precise positions of visual marks.
It supports nested hierarchies and configuration components, such as form elements mapping data attributes to axes.
Through metadata and accessibility tags (e.g., aria-attributes~\cite{nurthen2025aria}), it also \defrequirementBegin{reveals how visual variables are encoded and provides additional descriptions of data attributes} that a pure image cannot convey.
\textit{\textbf{Interaction Capabilities.}} The DOM includes event listeners (e.g., for mouse events) that define interactivity.
From these listeners, agents can \refrequirementDetails{02}{infer baseline interactions such as \texttt{click} and \texttt{hover}}.
Higher-level interactions, such as brushing—a sequence of mouse-down, mouse-move, and mouse-up—may also be \defrequirementBegin{inferred when tags have specific attributes or names}, though this is not guaranteed.
\textit{\textbf{Agent Accessibility.}} AI agents must \defrequirement{07}{Sync States}{access the DOM at runtime}
 and process its textual representation.

\subsection{\protect\reflayer{L3}~Declarative Grammar Representation}

\noindent\textit{\textbf{Definition and Scope.}} A declarative grammar describes the interface's format through rules or guidelines for creating consistent, effective visualizations.
Such grammars define a common vocabulary and structure for building and interpreting visualizations.
In the VA application design process, this is a low-level abstraction: everything is strictly specified by the template schema before compilation into layer \reflayer{L2}.
\textit{\textbf{Format and Structure.}} Common grammars such as Vega, Vega-Lite, and Mosaic's vgplot use \texttt{JSON} or \texttt{YAML}.
These definitions serve as blueprints for the interactive visualizations in the rendered application.
\textit{\textbf{Encoded Knowledge and Added Value.}} Based on these definitions, this layer encodes knowledge of attribute visualization, hierarchies, layouts, and rules for color selection and chart types.
It \defrequirementBegin{explicitly maps data attributes to visual objects.} Unlike the DOM, which represents the computed visual components, the declarative grammar \defrequirementBegin{captures the precise intent and configuration of the visualization design.}
\textit{\textbf{Interaction Capabilities.}} Interactions are explicitly defined and linked to specific data attributes and visual objects in the grammar.
Higher-level interactions like brushing are directly declared, listing available interactions.
Additional \refrequirementDetails{05}{semantic knowledge can be inferred from the interaction definitions and the data variables they manipulate.}
\textit{\textbf{Agent Accessibility.}} The declarative grammar-based definitions \refrequirementDetails{07}{must be injected into the runtime to be accessible to an AI agent} via specific tools or DOM access.
As with layer \reflayer{L2}, an AI agent needs text-parsing capabilities to interpret (semi-)structured text schemas.

\subsection{\protect\reflayer{L4}~Semantic Knowledge Representation}
\noindent\textit{\textbf{Definition and Scope.}} The production logic encodes semantic knowledge generated during VA design and development.
While layers \reflayer{L1} -- \reflayer{L3} describe composition and outcome, layer \reflayer{L4} captures the ``why'': design decisions, rationales, and intended purposes of specific interactions or visual marks, offering a holistic application view. %
\textit{\textbf{Format and Structure.}} As these details can be any kind of information, there is no fixed format or template.
When available, they are usually documented as free-text annotations by the developer, such as code comments.
\textit{\textbf{Encoded Knowledge and Added Value.}} This semantic knowledge reveals \defrequirementBegin{insights into the underlying data, machine learning model, and intended tasks or user groups.} It explains why a data attribute is encoded in a specific way and whether a VA application suits particular tasks.
\textit{\textbf{Interaction Capabilities.}} This layer's interaction knowledge is semantic rather than functional.
It \defrequirementBegin{describes metadata}, such as an interaction's purpose, instead of just technical event listeners or grammar definitions.
\textit{\textbf{Agent Accessibility.}} VA applications do not naturally expose this knowledge layer to users or agents, \defrequirementBegin{requiring extra logic or methods to access it at runtime.} Because the knowledge is mostly encoded in free text, AI agents \refrequirementDetails{06}{require robust text parsing and natural language understanding to interpret the semantic information}.

\section{VACP: the Visual Analytics Context Protocol}
\label{sec:VACP}

To bridge the gap between human-oriented VA interfaces and AI agents, there is need for a protocol that meets both their requirements and mediates between them. The VA application requires strict system logic and state consistency, while AI agents need semantic richness and a way to perceive and act on the VA interface. We, therefore, propose the \textbf{Visual Analytics Context Protocol (\texttt{\textsc{VACP}})}.

VACP serves as a formal abstraction layer that decouples the agent from implementation details (e.g., DOM structures or pixel data), while preserving a holistic semantic understanding of the analytical environment. It comprises three components: (1) \textbf{Application State Representation}, modeling the interface and its elements as a temporal semantic graph; (2) \textbf{Exposition of Available Interactions}, formalizing valid state transitions; and (3) \textbf{Interaction Execution Gateway}, translating the agent's intent into executed interactions. We ground our context protocol in principles that we consider the minimal requirements for enabling agents to navigate data-rich environments, even beyond VA applications. Some principles inform the agent through \refcontext{Contextualization}, while others ensure operations and \refdynamic{Interactive Dynamics} in the VA interface. In the following, we first introduce the principle, then explain its application in VACP and annotate the challenges (identified in \cref{sec:KnowledgeApplicationStateAccess}) that we address with the principle.

\subsection{Application State Representation}
\dimension{P1 -- App State Exposure}{VACP_color}
{To reason about an interface, the agent needs an up-to-date, semantically grounded representation of the application state.}
{\refcontext{Application State Contextualization}
}{Unlike experienced human VA users, who intuitively grasp visual metaphors, AI agents require explicit encoding of component purposes and the developer’s design decisions.
We encode the typical hierarchical structure of VA application in the state graph.
\textbf{Nodes} represent distinct VA components from global settings and dataset selection to fine-grained elements like interaction feedback. \textbf{Edges} capture structural containment (e.g., a ``Dashboard'' node containing a ``Scatterplot'' node) and semantic relations, such as coordinated views (e.g., ``Selection in Plot A filters Data in Plot B'').  
In VACP, agents access the (semantic) application state via \textit{get\_state()}, and layers~\reflayer{L1} and \reflayer{L2} via \textit{get\_chart()} and \textit{get\_dom\_extract()}. (\refrequirement{01},~\refrequirement{07})}

\dimension{P2 -- Metadata Annotation}{VACP_color}{\replace{By adding descriptive metadata about component purpose or behavior, we lower the reasoning barrier for agents.}{The reasoning barrier for agents can be lowered by adding descriptive metadata about component purpose or behavior.}}{\refcontext{Metadata Contextualization}}
{To support reliable reference and reasoning, the state representation adheres to the following constraints: \textbf{Unique Identification:} Each node has a stable, unique ID. So, the agent can consistently reference components across a session. \textbf{Temporal Consistency:} The application is treated as a living document. Each state snapshot includes a timestamp and unique snapshot ID to avoid race conditions and ensure the agent uses the latest interface state. \textbf{Semantic Annotation:} Nodes are annotated with metadata about component type, semantic layer, and a natural language description of their purpose. \replace{While optional}{Optional} metadata \add{further} enriches context and eases the agent’s reasoning. (\refrequirement{05})}

\dimension{P3 -- Agent Constraint Awareness}{VACP_color}{Considering agent constraints, such as context window size, optimizes agent runtime and reduces the risk of performance degradation.}{\refdynamic{Interactive Dynamics}
}{Crucially, VACP addresses the finite context window of LLM-based AI agents with a details-on-demand architecture.
Instead of serializing all data into state and hitting token limits, the state only exposes the data’s \textit{structure} and \textit{availability}.
The agent can use \textit{inspect\_data()} to retrieve only the specific data points relevant to its current task. (\refrequirement{05}, \refrequirement{06})}

\begin{figure*}[!tb]
  \centering
  \includegraphics[width=\textwidth, alt={Example VACP app-state graph where nodes represent VA entities, edges encode semantic relations, and state values are keyed by stable references.}]{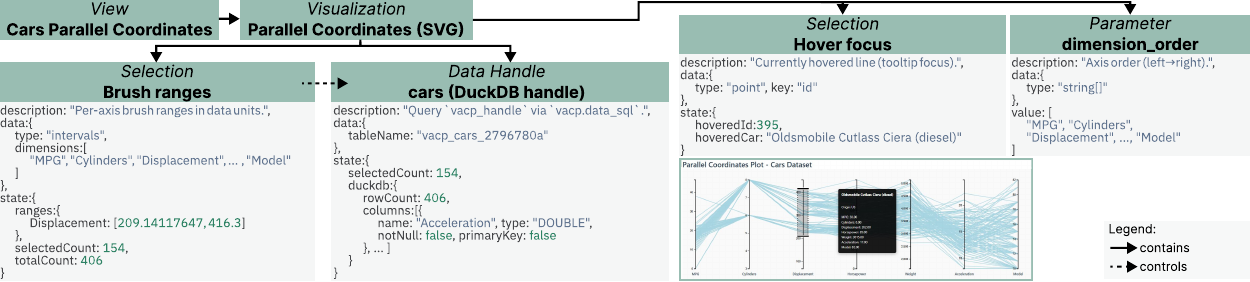}
  \vspace{-1.75em}
  \caption{Example VACP application-state representation. The capability graph captures semantic structure and interaction-relevant relations; the state snapshot stores currently active values under the same stable references.}
  \vspace{-1.5em}
  \label{fig:vacp_app_state}
\end{figure*}

\dimension{P4 -- Provenance Tracking}{VACP_color}{\remove{When interacting with the VA interfaces, agents might reach a state that is a dead end. Reverting to previous states might require multiple interactions. }Providing direct jumps back to a previous state improves the performance of the agent. Provenance tracking is optional.}{\refcontext{Temporal Contextualization}
}{%
\add{When interacting with the VA interfaces, agents might reach a dead end. Reverting to previous states might require multiple interactions. This can be overcome by allowing jumping to previous states.}
}
\subsection{Exposition of Available Interactions}

Interactive VA systems are defined by the dynamic relationship between the user and the system. In VACP, an \textbf{interaction} is an intent-driven agent action that triggers a system reaction, resulting in a mutation of the application state.
We formalize the state updates as follows. Let $V = \{v_1,\dots,v_n\}$ be the set of visual components and $I = \{i_1,\dots,i_m\}$ the set of available interactions. The application state at time $T$, $S_T$, is defined by the configuration of these components and available interactions:
$S_T = \{V_T, I_T\}$.
Applying an interaction $i_{x_T} \in I_T$ with parameters $P = \{p_1,\dots,p_k\}$ produces the next state:
$S_{T+1} = i_{x_T}(S_T, P)$.

\dimension{P5 -- Interaction Exposure}{VACP_color}{All currently valid interactions need to be explicitly exposed to the agent. This exposure is dynamic.}{\refcontext{Interaction Contextualization}
}{\replace{VACP requires registering available interactions provided to the agent by observing available interaction capabilities.}{To enable the agent to observe the available interaction capabilities, VACP requires registering available interactions at runtime.}
For safe execution, the protocol specifies interaction parameters, indicating which are required or optional, their data types, and valid ranges. Additional descriptions provide functional context. Agents can \replace{view}{query} all available interactions via \textit{get\_capabilities()}. (\refrequirement{02})}

\dimension{P6 -- Dynamic Interaction Update}{VACP_color}{
Upon an application state update, the protocol must dynamically update the list of capabilities to reflect the currently available interactions.   
}{\refdynamic{Update Dynamics}
}{To address the dynamics of the systems, VACP allows for dynamic registration or release of interactions to reflect the current context. Interactions are represented as \textbf{action nodes} within the state graph, linked to the components they manipulate. (\refrequirement{02})}

\dimension{P7 -- Tailored Data Access}{VACP_color}{Instead of serializing all underlying data into state, which would rapidly exhaust token limits, the state exposes only the \textit{structure} and \textit{availability} of data. The agent then queries specific data points only when they are relevant to the current task.}{\refcontext{Data Contextualization}
}{Additionally, VACP mitigates data-access bottlenecks with two actions:
\textit{get\_schema()}: Lets the agent inspect the data’s logical structure and constraints without loading records.
\textit{inspect\_data()}: Lets the agent run structured queries to retrieve subsets or aggregates of data tied to a visual component, delegating arithmetic to the system engine and avoiding LLM arithmetic calculation errors. (\refrequirement{06})}

\subsection{Gateway for Direct Interaction Execution}

\dimension{P8 -- Interaction Execution Gateway}{VACP_color}{To close the perception–action loop, the agent requires an API to execute interactions, giving it full accessibility beyond view-only mode and allowing it to manipulate the application state.}{\refdynamic{Execution Dynamics}
}
{The final VACP component is the execution gateway, which links perception to action and serves as the operational API for the agent’s action calls.
When \textit{execute\_interaction()} is invoked, the gateway runs multi-step validation: it resolves the interaction reference ID to the VA system function, verifies that the interaction is still valid for the current state snapshot (handling latency-related conflicts), and checks that the provided parameters satisfy the type and range constraints defined during exposition.
The gateway executes the interaction synchronously, immediately applying state updates to minimize \replace{latency}{execution time}. \replace{It then returns structured feedback to the agent:}{The agent receives structured feedback.}
On success: a confirmation status or, for data queries, the requested results.
On failure: a descriptive error message that aids agentic reasoning, helping diagnose issues and attempt self-correction.
The gateway can also support advanced mixed-initiative features such as interaction provenance logging, undo/redo stacks for safe exploration, and ``human-in-the-loop'' hooks that require user confirmation for high-stakes actions. (\refrequirement{02}, \refrequirement{03}, \refrequirement{04})}

\subsection{Agent Workflow using VACP}
\cref{fig:teaser} shows how AI agents \add{can} use VACP to interact with a VA interface. The agent first inspects the application state by interpreting the interface structure, VACP metadata, and interactive elements. It then chooses an action to advance its analysis, such as changing a visual data mapping, which updates the interface. The agent reads the updated state via VACP to verify the change. \replace{If it needs to}{To} focus on a data subset, it can brush the line chart. The system converts this into a selection, updates the application state, reflects the selection in the interface, and VACP passes this updated state back to the agent for its next step.

\section{The VACP Library}

We implement\replace{ VACP as a TypeScript library}{ a TypeScript library of VACP} for web-based VA systems\textsuperscript{\ref{foot:github-link}}. The library decouples the semantic interaction contract from the DOM layout, rendering backend, and UI framework. To integrate \add{VACP} with existing ecosystems, we provide adapters for declarative grammars such as Vega-Lite~\cite{satyanarayan2017vegaLite} and Mosaic's vgplot~\cite{heer2024mosaic}, \add{an API for custom solutions,} plus \replace{a Model Context Protocol~(MCP)}{an MCP} server for transport~\cite{Antropic_2025_MCP}. 

\vspace{.25em}
\iviaparagraph{Design Goals} The implementation targets four goals: \textit{Minimal setup} via default capabilities and state mappings; \textit{Direct integration} with common VA stacks to retrofit existing interfaces without changing interaction logic; \textit{Extensibility} through domain-specific metadata and custom actions; and \textit{State consistency} by synchronizing a single application state across human and agent interactions.

\subsection{Semantic Grammar: State and Interactions}
The architecture isolates protocol semantics from transport layers. At runtime, the application defines three synchronized objects: a capability graph, a state snapshot, and an action catalog. Transport layers pass requests to this runtime without changing the\remove{ir} semantic meaning.

\vspace{.25em}
\iviaparagraph{State Representation}
The library divides application state into structural capabilities and \replace{live}{volatile} values \remove{as shown in }(\cref{fig:vacp_app_state}). Structural capabilities are mostly static and define the semantic space via the capability graph. \replace{Live}{Volatile} values change during interaction and encode current selections or control settings in the state snapshot. Both layers use stable semantic references, letting agents track entities across multi-step interactions\remove{ without re-identifying DOM elements after each update}. Developers can optionally attach metadata\remove{such as titles and summaries to these references} to clarify domain semantics when multiple valid operations exist.

\vspace{.25em}
\iviaparagraph{Action and Data Representation}
VACP models interactions as semantic actions instead of low-level UI events. Each catalog action has a stable identifier, parameter schema, and optional target reference. \remove{For data access,} VACP offers structured queries on referenced data handles, letting agents retrieve relevant datapoints and aggregates without reading from pixels or the DOM. This bounds data extraction and avoids serializing full datasets into the LLM context.

\begin{figure}[!b]
  \vspace{-1.5em}
  \centering 
  \includegraphics[width=\columnwidth, alt={An overview of the evaluation pipeline. Left (a) shows the overall task execution and evaluation workflow, and right (b) illustrates the agent setup and the different evaluated scenarios S1-S4.}]{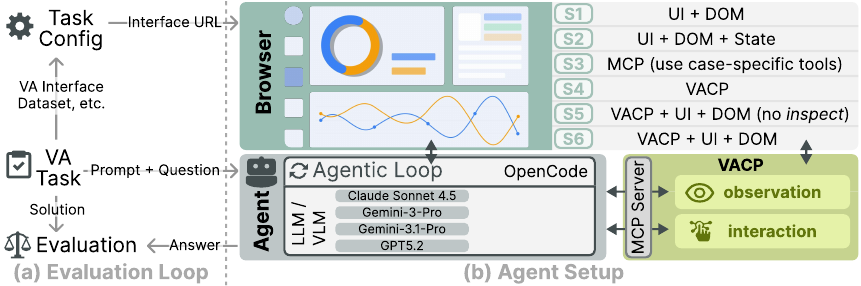}
  \vspace{-1.75em}
  \caption{%
  	\showRevisionUpdate{[Figure updated] }Overview of the Evaluation pipeline, where the left (a) shows the overall task execution and evaluation workflow, and the right (b) illustrates the agent setup and the different evaluated scenarios \protect\reflmprinciple{S1}~--~\protect\reflmprinciple{S6}.  
  }
  \label{fig:evaluation_pipeline}
\end{figure}

\subsection{Execution and Transport Loop}
Interaction execution uses a validated request–response cycle. When an agent submits an action, the runtime validates parameters, runs the framework logic, and returns structured feedback. This explicit channel surfaces execution errors for self-correction. Afterward, the agent queries the updated state via stable references to verify the outcome.

To reduce LLM context limits and \replace{latency}{execution time}, the implementation avoids continuous full-state serialization, scoping state updates to semantic references and returning them as incremental diffs.

\vspace{.25em}
\iviaparagraph{MCP Agent Accessibility}
We provide an MCP server that acts exclusively as a transport layer for VACP, \add{exposing a small, static set of core tools for state observation and action execution}. Since updating the MCP tool registry after each UI change would be too costly in tokens and \replace{latency}{execution time}, we deliberately avoid dynamically mapping VA interactions to MCP tools.\remove{Instead, the MCP server exposes a small, static set of core tools for state observation and action execution.} The agent uses the tools to query the internal VACP action catalog and decides when to refresh its view of available interactions. If it attempts an obsolete action, the VACP runtime returns a structured error, prompting the agent to adjust.

\subsection{Integration Examples}
The library uses adapter patterns to map framework-specific interactions to the shared VACP contract, supporting two common VA paradigms: declarative grammars and custom imperative visualizations. In \cref{tab:dev_effort} we show the implementation effort for our use cases.

\textbf{Declarative libraries}: an adapter parses the specification to extract encodings, parameters, and selections. A wrapper maps native Vega-Lite specifications to protocol actions while preserving execution logic.

\begin{figure}[!h]  \vspace{-1em}
  \centering
  \includegraphics[width=\linewidth]{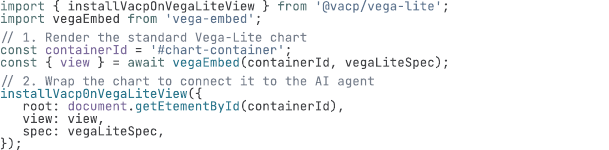}
  \vspace{-2em}
  \label{fig:code_vacp_vl}
\end{figure}

\textbf{Custom charts}: without a standardized internal model, developers manually build the semantic bridge by defining stable references to interactive components, exposing framework-specific functions as semantic actions, and adding a polling function to sync live state.
\begin{figure}[!h]   \vspace{-1em}
  \centering
  \includegraphics[width=\linewidth]{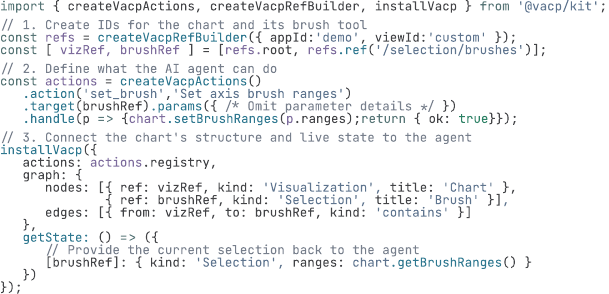}
  \vspace{-2em}
  \label{fig:code_vacp_d3}
\end{figure}

\begin{table}[!b]
   \vspace{-1.0em}
  \caption{%
  	\add{Required lines of code (LoC) for adding VACP to use cases.}%
  }
  \vspace{-1em}
  \label{tab:dev_effort}
  \scriptsize%
  \centering%
  \newcommand{\tightheader}[1]{\linespread{0.75}\selectfont #1}
  
  \begin{tabularx}{\linewidth}{Xp{2cm}*{3}{m{0.9cm}}}
  	\toprule
  	\tightheader{\add{use cases}} & 
    \tightheader{\add{framework}} & 
    \tightheader{\add{total\newline (LoC)}} & 
    \tightheader{\add{semantic\newline (LoC)}} & 
    \tightheader{\add{action\newline (LoC)}}\\
  	\midrule
  	\add{UC1, UC2, UC4} & \add{Vega-Lite} &     \add{37}  & \add{0} & \add{0} \\
  	\add{UC3} & \add{Mosaic vgplot} & \add{42}  & \add{0} & \add{0} \\
  	\add{UC5} & \add{React + D3} &    \add{395} & \add{47} & \add{224} \\
  	\bottomrule
  \end{tabularx}%
\end{table}

\section{Evaluation}
\label{sec:evaluation}
We evaluate VACP-enabled agents against state-of-the-art approaches to validate the VACP framework’s effectiveness and improvements. We detail our methodology, evaluation setup, and findings.

\subsection{Methodology}

To systematically evaluate and improve AI web agents in VA, we established a multi-stage process encompassing the curation of a proof-of-concept benchmark dataset, expert task validation, identification of agent limitations, and the development and evaluation of VACP.

\vspace{.25em}
\iviaparagraph{Benchmark Development and Task Formalization}
While visualization literacy tests like VLAT~\cite{lee2017VLAT} exist, to the best of our knowledge, there is no standardized benchmark \remove{dataset }for assessing users' ability to use interactive VA systems to solve problems. \replace{To fill this gap, we curated a representative set of baseline VA systems and tasks.}{Addressing this, we curated a task set of baseline VA systems and tasks from established literature, such as visualization design spaces and literacy tests, that experts validated as individually indicative of prominent domain practices.} \remove{Our goal was to test whether AI web agents can perform fundamental VA interaction primitives, prioritizing atomic tasks rather than complex workflows to isolate specific perception and execution failures.}\add{We prioritized atomic tasks rather than complex workflows to isolate specific perception and execution failures.} 

\replace{We identified a representative set of tasks and systems from established literature, such as visualization design spaces and literacy tests. Because real-world VA applications use diverse web technologies, our benchmark needs to reflect this diversity. We therefore implemented the test systems using two distinct approaches: custom, low-level code (React and D3.js) and high-level specification languages (Vega, Vega-Lite, and Mosaic vgplot).}{To reflect on the diversity of web technologies in real-world VA systems, we implemented the test systems using two distinct approaches: custom, low-level code (React and D3.js) and high-level specification languages (Vega, Vega-Lite, and Mosaic vgplot).} This variety allows testing whether an agent can adapt to different web structures and levels of semantic complexity. 
The final tasks \replace{and systems were}{suite (see \mbox{\cref{sec:UseCases}}) was} refined through iterative expert discussions among the authors.\remove{Further details on these use cases are provided in \mbox{\cref{sec:UseCases}}.}
\remove{After selecting the use cases, we implemented the tasks across the identified implementation spaces, resulting in five use cases: one React-based (Parallel Coordinates Plot), three Vega-Lite, and one vgplot example.}
\begin{figure}[b!]
  \vspace{-1.75em}
  \centering 
  
  \includegraphics[width=\columnwidth, alt={Overview of VA interfaces of the defined VA use cases. Use case id and the visual interface are provided for each use case.}]{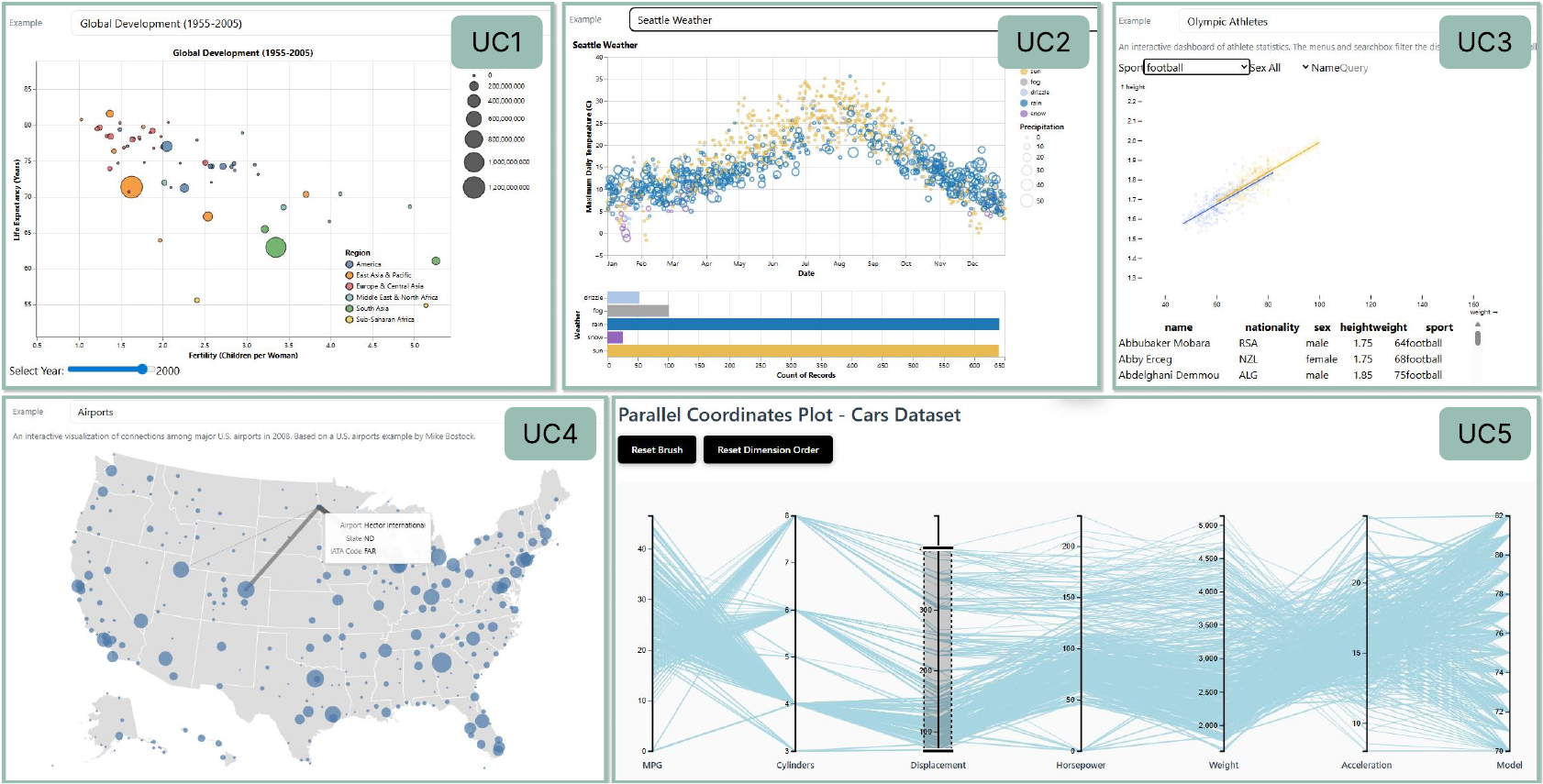}
  \vspace{-1.75em}
  \caption{%
  	Overview of VA interfaces of the defined VA use cases.
  }
  \label{fig:usecase_overview}
\end{figure}
\vspace{.25em}
\\\iviaparagraph{Pre-Study and Expert Validation}
To ensure our benchmark was solvable and \replace{representative}{represents real VA scenarios}, we conducted a formative pre-study with VA experts, who interacted with the selected VA systems to solve the defined tasks.
The study aimed to (1) verify task solvability, (2) assess the suitability of the visualizations and interactions, and (3) evaluate the representativeness of system components. It yielded critical feedback for optimizing our experimental design. Experts confirmed that the tasks, visualizations, and interactions \replace{were representative}{represent real VA challenges} but identified ambiguities in several task descriptions. We refined these descriptions to ensure that subsequent agent failures could be attributed to functional limitations rather than prompt ambiguity.

\vspace{.25em}
\iviaparagraph{Baseline Agent Evaluation}
We evaluated current state-of-the-art web AI agents on our benchmark and found that they fail most tasks (completion rate 28\%–\replace{51}{73}\%). \replace{Our analysis shows that agents}{Agents} miss the visualization’s semantic context, struggle to access data details hidden within the DOM, and lack the precision for accurate interactions.

\vspace{.25em}
\iviaparagraph{Comparative Evaluation}  
We evaluated the efficacy of our approach by implementing VACP and giving AI agents access to different context scenarios. This summative evaluation tested whether agents using VACP could complete the analytic tasks. We compared VACP against alternative context types\remove{ (e.g., raw UI and DOM access)}, measuring task completion, duration, and token usage.
For deeper analysis, we annotated each agent turn with dimensions based on foundational HCI literature~\cite{buxton1983inputStructures,norman2002design,pirolli2005sensemaking} to quantify agent behavior, recognizing that solution paths are highly iterative and that successes and failures are not random prompt artifacts, but rather structural consequences of navigating or failing to navigate the VA pipeline based on the provided knowledge and interaction access.

\begin{figure*}[htb!]
  \vspace{-2em}
  \centering 
  \includegraphics[width=\textwidth, alt={Workflow of Claude Sonnet 4.5 successfully solving Task UC1L in scenario. After inspecting a screenshot and the semantic state information provided by VACP, the agent first adjusts the selected year, then verifies the change by observing the UI. Based on querying the provided data, the agent identifies ``Japan'' and verifies the result by executing a hover on the data point of Japan. After the verification, the agent finally returns ``Japan'' as the correct response.}]{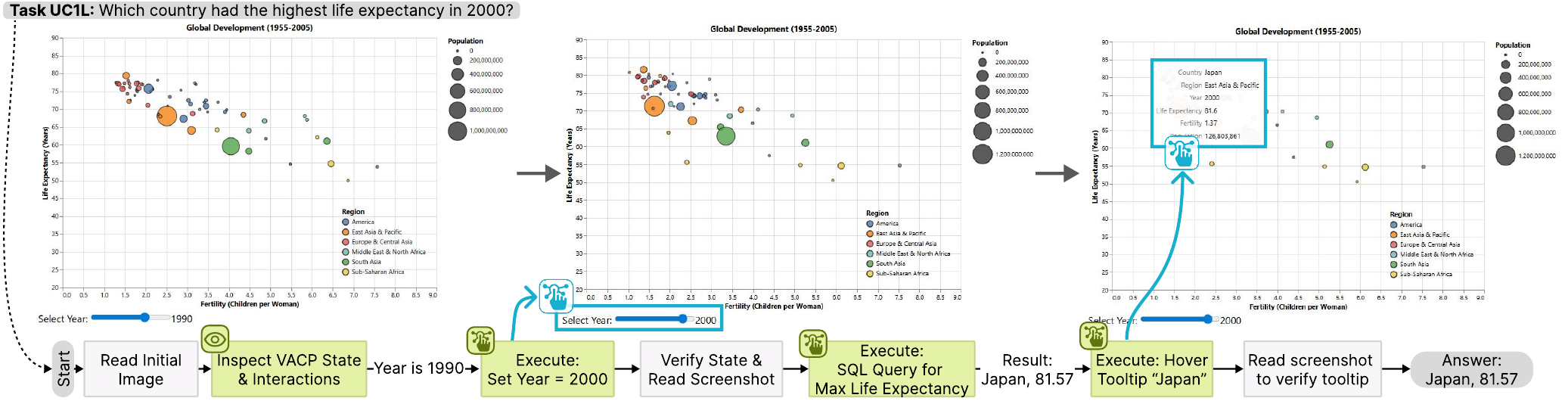}
  \vspace{-1.75em}
  \caption{%
  Claude Sonnet 4.5 solving Task UC1L in \protect\reflmprinciple{S6}. After inspecting a screenshot and the VACP semantic state, the agent adjusts the selected year and verifies the change in the UI. It then queries the data, identifies and confirms ``Japan,'' by hovering the data point, and returns the correct answer.   
  }
  \vspace{-1.5em}
  \label{fig:exampleWorkflow}
\end{figure*}

\subsection{Use Case and Task Description} \label{sec:UseCases}

To rigorously evaluate AI agents in VA, we used the previous curated set of \replace{representative}{compelling} use cases spanning a comprehensive design space. The scenarios cover diverse data modalities, visualization techniques, and complex interactions \add{from existing taxonomies~\cite{munzner2014analysisAndDesign,brehmer2013typology,bach2023dashboard}. The use cases span all core data types from Munzner~\cite{munzner2014analysisAndDesign} instantiated within a targeted visual representation space of numbers, tables, and detailed visualizations~\cite{bach2023dashboard}. Available interactions include select, navigate, arrange, change, and filter, excluding aggregation, introduction, and dashboard-reconfiguring encoding tasks (see~\cite{brehmer2013typology}, How).} To probe technical limits, we include both declarative grammars and imperative paradigms (React, D3.js), requiring agents to manipulate SVG-based DOM elements and pixel-based canvas renderings (see~\cref{fig:usecase_overview}).

\noindent\textbf{UC1 -- Global Development:} This interactive bubble chart visualizes health demographics over time. It supports temporal animation via sliders, tooltips for details, and geometric zooming/panning to resolve occlusion in dense regions.
\textbf{UC2 -- Seattle Weather:} A coordinated multiple-view system links a temperature scatterplot with a weather-frequency bar chart. The tightly coupled views support bidirectional cross-filtering—temporal brushing updates distributions, categorical selection filters records, and hover reveals precise details.
%
%
\textbf{UC3 -- Olympic Athletes:} A dashboard combining a biometric scatterplot (with gender-based regression lines) and a sortable data table, supporting complex queries via spatial brushing and filters for sport, gender, and keywords.
%
%
\textbf{UC4 -- U.S. Flights:} A geospatial node-link diagram overlaid on a map supports egocentric network analysis. Clicking an airport node filters outgoing flight edges, while hovering reveals weighted edge attributes (flight counts) and node metadata.
\textbf{UC5 -- Technical Car Specs:} A Parallel Coordinates Plot for high-dimensional automotive data~\cite{quinlan1993cardataset}. It enables multivariate analysis through multi-axis brushing for concurrent constraints, drag-and-drop axis reordering to expose correlations, and global reset controls.

\vspace{.25em}
\iviaparagraph{Task Definitions} \remove{We defined the tasks based on existing VA task frameworks and typologies}%
\remove{to obtain a representative set of analytical tasks.} 
\add{We define the tasks using Brehmer and Munzner’s multi-level typology~\cite{brehmer2013typology}. From the Why dimension, our evaluation targets locate, identify, and compare actions, excluding produce and summarize. Agents must use latent lookup, browse, and explore behaviors as primitives to obtain correct answers.} For each use case UCx, we specify three tasks (UCxL, UCxI, UCxC) that differ in required interactions and focus on elementary analytical tasks: \textit{locate} (UCxL), \textit{identify} (UCxI), and \textit{compare} (UCxC). Some tasks involve multiple elementary actions, but the primary one is indicated by the index. \add{Ground truth solutions are derived from the underlying datasets and results are validated via semantic similarity (textual values) and numeric tolerances (numerical values) defined in the evaluation setup.} All task descriptions, \add{mappings to taxonomies,} solutions, and the validation setup are provided in the supplemental material.

\subsection{\replace{Verification of Task \& Interface Representativeness}{Validation of Tasks \& Interfaces}}
\label{sec:expertStudy}

To align our benchmark with real VA challenges, we conducted an expert user study to verify that our use cases and tasks form a representative baseline for evaluating AI agents.

\vspace{.25em}
\iviaparagraph{Study Design and Demographics}
We recruited 7 participants~(Pt) \add{(six domain experts in VA and HCI, and one non-expert)} to complete 15 tasks across our VA use cases \add{ and to assess expert alignment and interface intuitiveness.} \remove{: six domain experts in VA and HCI, and one non-expert. This enabled us to assess expert alignment and interface intuitiveness.}%
Participants rated cognitive demand and tediousness, and judged task representativeness on a 5-point Likert scale.

\vspace{.25em}
\iviaparagraph{Representativeness of Interactions and Tasks}
Participants rated the interactions as highly representative (Median $= 5, \mu \approx 4.6$). Pt4 described them as ``very natural interactions,'' mirroring how a human analyst explores data. Pt1, Pt2, Pt5, Pt6, and Pt7 agreed that ``all interactions made sense.'' Participants rated the tasks as highly representative (Median $= 4, \mu \approx 4.4$). Pt4 noted the tasks cover a wide range, including diverse datasets, variations, temporal data, and geospatial patterns. Pt1 said real VA tasks are ``more complex and open-ended'' than the atomic tasks, but we emphasize that these are the basic building blocks of analysis.\remove{ If } AI agents \replace{struggle}{struggling} with \remove{the }foundational tasks\remove{, they} are not ready for the ambiguity of real-world exploration.

\vspace{.25em}
\iviaparagraph{Task Correctness and Ambiguity Resolution}
Human participants achieved an 85.71\% completion rate. No task stumped all participants, confirming solvability. The study revealed specific challenges that informed our final task descriptions: \textit{Visual Complexity:} UC4 is difficult because dense datapoints create overlapping lines, obscuring routes. This shows the testbed captures real-world ``messiness,'' where occlusion challenges agents. \textit{Ambiguity:} Pt2, Pt4, Pt5, and Pt6 flagged ambiguous terms in \replace{Global Development}{UC1} (finding the ``largest'' country) and \replace{U.S. Flights}{UC4} (defining ``smallest'' airport) use cases. We revised prompts to remove ambiguity while keeping visual difficulty. \textit{Precision vs. Estimation:} Accuracy varied by strategy: participants using hovers and tooltips were more precise than those visually estimating from axes. This distinction is crucial for evaluation, separating agents that can ``read'' the DOM from those that must ``see'' the rasterized chart.

\vspace{.25em}
\iviaparagraph{Interface Design and Usability}
Including a non-expert (Pt5) helped test usability. Despite lacking VA expertise, Pt5 identified interactions and completed tasks, indicating intuitive interfaces. Pt3 felt that some visual encodings could be improved, but said that such imperfections are common in real-world interfaces. 
Pt5 requested ``more precise controls, like entering explicit numbers or text for filtering,'' supporting our VACP contribution, which gives agents more precise control options than standard interfaces.

\vspace{.25em}
\iviaparagraph{Summary}
This study shows our testbed is realistic \remove{and high-fidelity }for VA agent testing, with natural interactions and tasks that accurately reflect real visual and cognitive challenges in VA workflows.

\subsection{Agent Evaluation}
To validate the VACP framework, we evaluated AI agent performance under varying levels of contextual access, examining how different interface representations (from standard visual scene captures to our semantically enriched VACP interface) affect VA task completion.

\subsubsection{Evaluation Setup}
We utilize multiple AI agents composed of multimodal foundation models for text parsing, vision, and tool use, operating within a standardized agentic loop. This loop follows the ReAct~\cite{yao2022react} framework, allowing agents to iteratively reason, act, and observe feedback. For evaluation, we used human pre-validated VA tasks and interfaces (see \cref{sec:expertStudy}).

\vspace{.25em}
\iviaparagraph{Models and Sampling Strategy} To avoid artifacts tied to a specific model, we evaluate four SOTA models: \textit{GPT-5.2} (OpenAI)~\cite{openAI2025GPT52}, \textit{Claude 4.5 Sonnet} (Anthropic)~\cite{Anthropic2025ClaudeSonnet45}, \textit{Gemini 3 Pro Preview} and \textit{Gemini 3.1 Pro} (Google)~\cite{gemini2023geminiFamily}. Due to the non-deterministic nature of LLMs (e.g., temperature), we ran all experiments three times per model to capture variance in success rate, token usage, and runtime. This allows us to assess generalizability and identify cross-model behavioral patterns across scenarios. While runs differed in their reasoning paths, they produced similar outcomes, with only minor variation in token consumption and duration. \add{We added \textit{Gemini 3.1 Pro} to our evaluation as \textit{Gemini 3 Pro} was deprecated when running ablations \reflmprinciple{S3} and \reflmprinciple{S5}. 
}

\vspace{.25em}
\iviaparagraph{Scenarios} We evaluate agents in \replace{four}{six} scenarios, reflecting increasing context, knowledge representation, and interaction access:
\\\deflmprinciple{S1}~\textbf{UI + DOM:} Mirroring current general-purpose web agents, agents see only screenshots (\reflayer{L1}) and raw DOM (\reflayer{L2}). Complex visualizations (Vega-Lite/Mosaic) are rendered as canvas elements, so their data is hidden from DOM parsers. Only the PCP use case uses SVG. Agents act via emulated mouse events and JavaScript execution.
\\\deflmprinciple{S2}~\textbf{UI + DOM + State:} Building on \reflayer{L1} and \reflayer{L2}, we inject the instantaneous VACP application state into the DOM as serialized JSON in a hidden node. Agents can \add{discover and} access it via DOM queries.\remove{, but must first discover and parse this node.}
\\\add{\deflmprinciple{S3}~\textbf{MCP (with \textit{inspect}):} To compare VACP from simpler structured tool access, in this ablation, agents interact \textit{exclusively} via use case-specific static tools from an MCP server. While providing the state, interactions, and \textit{inspect\_data()}, there is no visual and DOM access.}
\\\deflmprinciple{S4}~\textbf{VACP (State + Interaction Gateway):} This ablation removes visual and DOM access. Agents interact \textit{exclusively} via tools from the VACP MCP server (\reflayer{L3}, \reflayer{L4}), relying entirely on the semantic API and testing the VACP schema’s expressiveness without visual grounding.
\\\add{\deflmprinciple{S5}~\textbf{VACP + UI + DOM (no \textit{inspect}):} Agents receive the full toolset (\reflayer{L1}~--~\reflayer{L4}), except \textit{inspect\_data}): This lets us study the contribution of interaction grounding and data access through \textit{incpect\_data()}.}
\\\deflmprinciple{S6}~\textbf{VACP + UI + DOM:} Agents receive the full toolset (\reflayer{L1}~--~\reflayer{L4}): VACP tools, DOM access, and screenshots, letting us study synergies or conflicts between visual perception and semantic data retrieval.

\vspace{.25em}
\iviaparagraph{Technical Implementation} We use OpenCode~\cite{opencode_2026} to manage the agentic loop, ensuring a robust, reproducible tool-execution environment. Each experiment runs in an isolated, sandboxed browser instance. \cref{fig:evaluation_pipeline} shows the evaluation setup. All agent runs \add{are limited to a max duration of 20 minutes without allowing for retries and} share the same system prompt, which specifies the task, expected response format (see supplemental material), and the initial interface URL. The system prompt was not updated during the runs. \add{We initialize all models with temperature 1.0 (recommended by the model providers).} In scenarios \reflmprinciple{S4}, \add{\reflmprinciple{S5}} and \reflmprinciple{S6}, agents access VACP via an MCP server exposing the available tools.
All experiments were run on an Apple MacBook Pro with an M3 Max chip and 36GB memory, using model providers’ APIs for inference. The evaluation source code \add{and details on API and models} \replace{is}{are} provided in the supplemental material.
\begin{figure}[t]
  \centering 
  \vspace{-.5em}
  \includegraphics[width=\columnwidth, alt={Summary of the AI agent evaluation on the provided use cases and tasks. Each tile presents the number of successful runs over three runs. Percentages after the agent model name indicate the overall success rate on the provided scenarios S1 -- S4, use cases, and tasks. VACP-enabled agents reach higher completion rates.}]{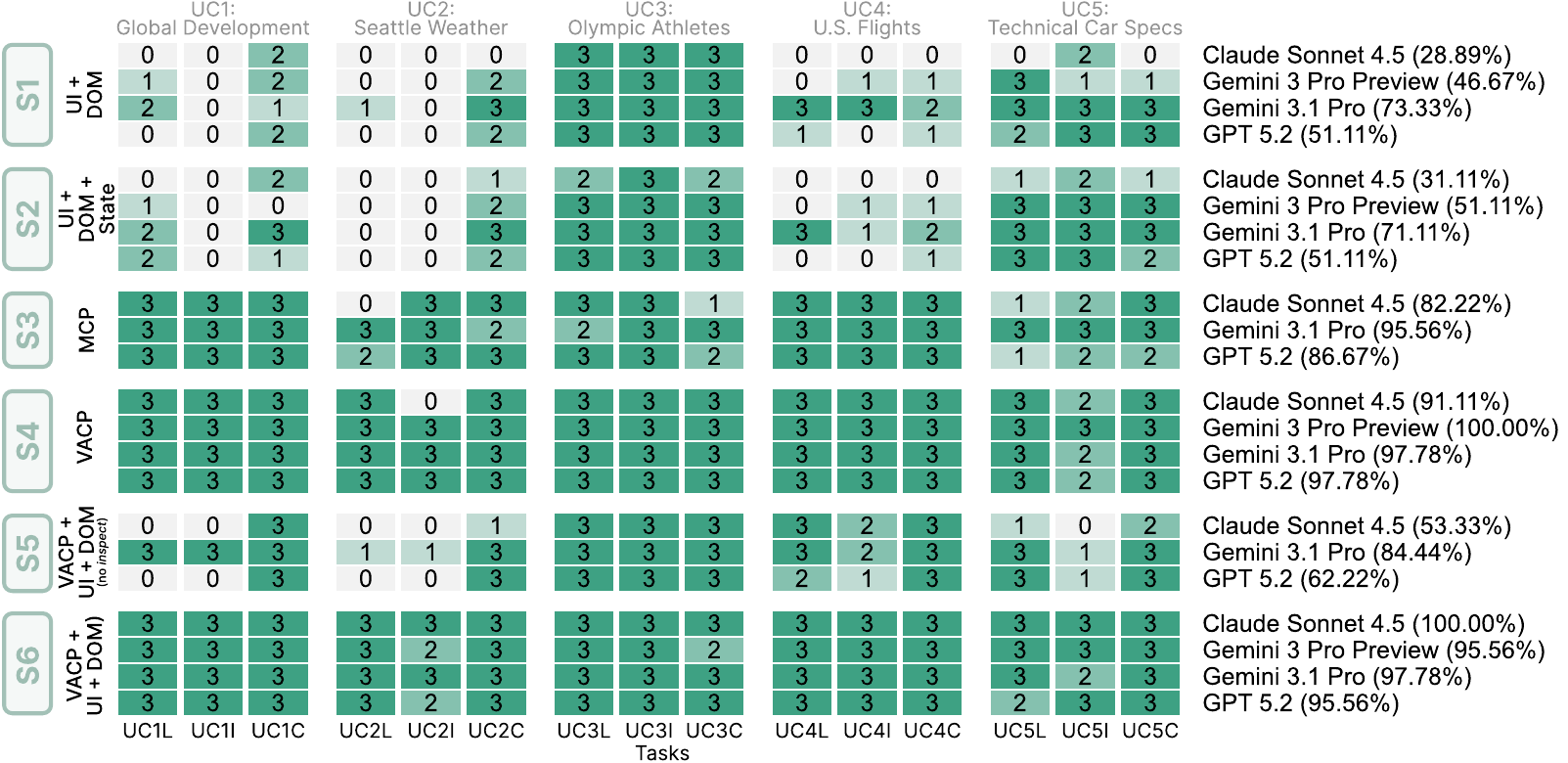}
  \vspace{-2em}
  \caption{%
  \showRevisionUpdate{[Figure updated] }Summary of the AI agent evaluation on the provided use cases and tasks. Each tile presents the number of successful runs over three runs. Percentages after the agent model name indicate overall success rate on the provided scenarios \protect\reflmprinciple{S1} -- \protect\reflmprinciple{S6}, use cases, and tasks. 
  }
  \vspace{-1.5em}
  \label{fig:evaluation_summary}
\end{figure}

\vspace{.25em}
\iviaparagraph{Initial Test Runs}
Prior to the main evaluation, we conducted pilot studies with a standard ``computer use'' loop~\cite{geminiComputerusePreview2026github}. Performance was suboptimal: agents struggled with controls such as dropdowns and inputs, and the setup did not support non-Gemini models. This resulted in the shift to the more robust, well-tested, open-source OpenCode framework. We also found that agents exploited ``loopholes'' instead of solving VA tasks, e.g., hallucinating answers from pre-training on standard datasets, downloading data from external GitHub repositories, or writing Python scripts to parse the DOM and bypass visual interaction. In response, we refined the system prompt to enforce strict tool use and restricted sandbox network access to prevent data leakage.

\subsubsection{Evaluation Results}
\label{sec:evaluation_res}
\cref{fig:evaluation_summary} shows our experiment summaries. Detailed traces of agents' perception, chain-of-thought, and actions are in supplemental material.

\vspace{.25em}
\iviaparagraph{Success Rates and the Visual Gap} We observe a performance gap between visual-based and semantic-based agents. Scenario \reflmprinciple{S1} yielded the lowest success rates (\replace{28\%--51\%}{28\%--73\%}) across all models: agents struggled to extract precise values from pixel-based canvas visualizations and DOM representations, confirming that current multimodal models are inadequate for precision-critical VA tasks. Adding the state to the DOM (\reflmprinciple{S2}) slightly improved success (\replace{31\%--51\%}{31\%--71\%}, max. gain 4.4\%). In contrast, scenarios relying on VACP's semantic interface (\replace{\reflmprinciple{S3}, \reflmprinciple{S4}}{\reflmprinciple{S4}, \reflmprinciple{S6}}) achieved near-perfect success across tasks. \add{Analysis of error types (\cref{appendix:residualerrors}) shows that in \reflmprinciple{S1} and \reflmprinciple{S2} the main failures stem from agents not fully exploring relevant data or giving up after several interaction attempts. Agents with structured tool access (\reflmprinciple{S3}) fail by misinspecting data, encountering execution timeouts, and not completing required interactions. In \reflmprinciple{S4} and \reflmprinciple{S6}, failed runs mostly result from misinterpreting or incorrectly requesting data. While succeeding in \reflmprinciple{S4}, one agent returned an unparsable response in \reflmprinciple{S6}, and another failed while deciding to reverse-engineer data from the DOM. Without \textit{inspect\_data} (\reflmprinciple{S5}), most failures are visual errors similar to those in \reflmprinciple{S1} and \reflmprinciple{S2}.}
Thus, providing a semantic interface significantly improves success rates, indicating that structured semantic state representations and API-driven interaction are more reliable for analytical reasoning and interaction than purely visual or textual DOM interpretation.

\begin{figure}[b]
  \centering 
  \vspace{-1.75em}
  \includegraphics[width=\columnwidth, alt={Summary of the AI agent evaluation. Detailed analysis of median token consumption and execution time for the different context scenarios, averaged on the use cases and evaluated models. VACP-enabled agents have lower token usage and shorter execution times. In addition, Claude consumes fewer tokens than other models.}]{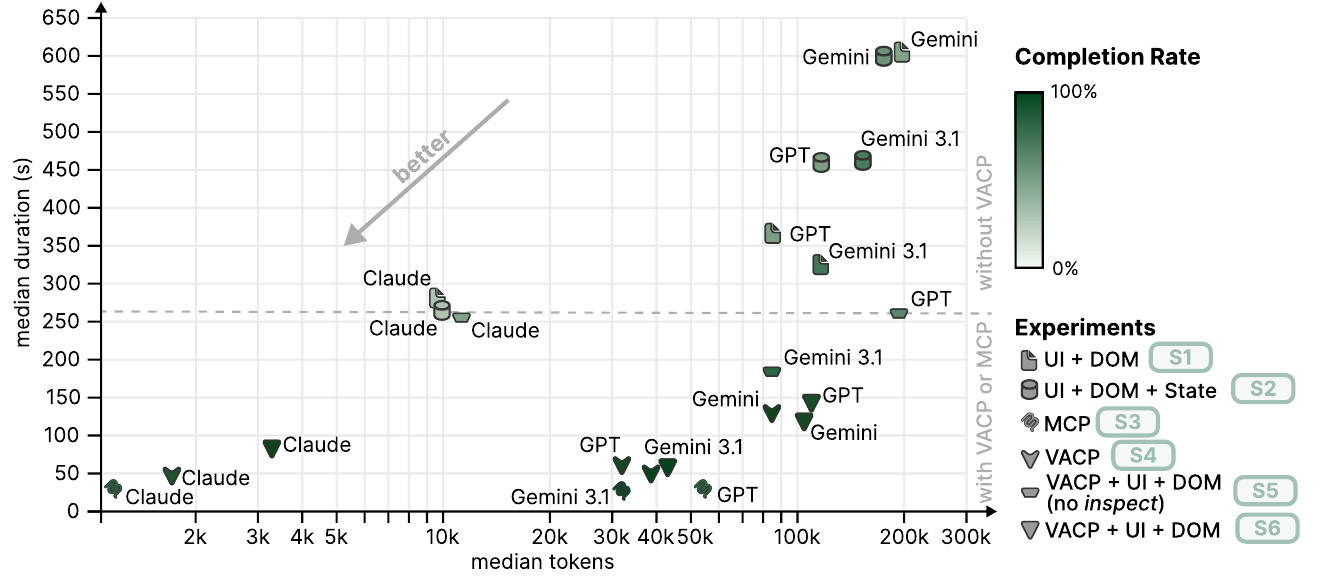}
  \vspace{-1.75em}
  \caption{%
  \showRevisionUpdate{[Figure updated] }Summary of the AI agent evaluation. Detailed analysis of median token consumption and \add{task} execution time for the different context scenarios, averaged on the use cases and evaluated models. 
  }
  \vspace{-1em}
  \label{fig:evaluation_tokenconsumption_duration}
\end{figure}

\vspace{.25em}
\iviaparagraph{Efficiency and Token Consumption} Interaction modality also affects computational efficiency. Agents in scenario \reflmprinciple{S4} (VACP) consumed vastly fewer tokens and finished faster than those in \reflmprinciple{S1}, \reflmprinciple{S2}, \replace{and \reflmprinciple{S4}}{\reflmprinciple{S5}, and \reflmprinciple{S6}} (\cref{fig:evaluation_tokenconsumption_duration}). \add{Agents provided with use case-specific tools (\reflmprinciple{S3}) consume the fewest tokens but have lower success rates.}
Removing DOM parsing and high-dimensional image processing allowed for leaner reasoning traces and interactions. The hybrid scenario \replace{\reflmprinciple{S4}}{\reflmprinciple{S6}} enabled ``visual verification'' (double-checking actions against screenshots), which consistently increased token usage and time on task, without a proportional increase in success rate. \textit{Claude Sonnet 4.5} agents consumed far fewer tokens on average (<2000 tokens for \replace{\reflmprinciple{S3}}{\reflmprinciple{S4}}) than ones with other models (e.g., >30000 tokens for \replace{\reflmprinciple{S3}}{\reflmprinciple{S4}} on GPT-5.2). 

\begin{addblock}
\vspace{.25em}
\iviaparagraph{Added Value of Structured Tool Access and Data Inspection} While consuming slightly more tokens (\cref{fig:evaluation_tokenconsumption_duration}), VACP provides value over simpler use case tailored structured tool access by achieving higher task completion rates (\reflmprinciple{S4}: 97\%, \reflmprinciple{S3}: 89\%). Further, in \reflmprinciple{S3} for two of the five ``locate'' tasks, where the visualizations are cluttered, the agents fail to visually perceive the values, as they estimate the values or use wrong data points as reference. In ablation \reflmprinciple{S5} we study the impact of direct data access. When restricting the data inspection (\reflmprinciple{S5}), the interaction grounding in the additional semantic metadata leads to higher completion rates than for the baseline (\reflmprinciple{S1}) and state injection (\reflmprinciple{S2}), but has lower success rates on ``identify'' tasks than on the other tasks, as they fail to read the exact values. Comparing \reflmprinciple{S5} and \reflmprinciple{S6}, agents with tools to inspect the data have improved completion rates while reducing the token consumption and execution time.  
\end{addblock}

\vspace{.25em}
\iviaparagraph{Agent Behavior and Interaction Stata/Analytic Provenance} While efficiency metrics highlight the computational benefits of semantic interfaces, explaining these performance gaps requires examining agents' iterative problem-solving strategies. To confirm that success rates stem from the provided context rather than prompt artifacts, we annotated every agent turn across four dimensions (\textit{Stratum}, \textit{Phase}, \textit{Outcome}, \textit{Adaptation}) grounded in foundational HCI literature~\cite{buxton1983inputStructures,norman2002design,pirolli2005sensemaking}.

We operationalize these human-centric frameworks for LLM agents by mapping Norman's Gulf of Execution~\cite{norman2002design} to an agent's ability to formulate valid API calls or DOM interactions, and the Gulf of Evaluation to its capacity to parse returned state representations or pixels. We similarly adapt Pirolli and Card's sensemaking loop~\cite{pirolli2005sensemaking} to characterize how agents update their internal context windows from intermediate results. For each turn in an agent run, we annotate where the agent is operating (\textit{Stratum}), the turn's intent (\textit{Phase}), whether the \textit{Outcome} was successful, and if the turn triggered a strategic \textit{Adaptation}. Annotations were generated with a structured LLM approach. To ensure methodological rigor, the authors manually validated on a random 20\% sample, achieving strong inter-rater reliability (Cohen's $\kappa = 0.918$\add{, between the human annotator and the LLM}). \add{In case of disagreement, the annotation of the LLM was corrected.} Because this work focuses on interface accessibility, we detail only the interaction stratum below. Full reports on intent, outcome, etc. appear in the supplemental material.

\vspace{.25em}
\iviaparagraph{Interaction Strata and the Verification Loop}
Following Buxton's taxonomy~\cite{buxton1983inputStructures}, we grouped interactions into three strata: pragmatic (manipulating raw pixels or coordinate-based events), syntactic (navigating the logical DOM or view structure), and semantic (interacting with raw data models, schemas, or analytical APIs).
Their distribution shows a strong link between semantic access and task success. In \reflmprinciple{S1}, agents relied almost entirely on syntactic (avg. 48.44\%) and pragmatic (avg. 47.19\%) interactions, with minimal semantics (avg. 4.37\%). Adding application state to the DOM in \reflmprinciple{S2} slightly increased semantics (avg. 11.82\%) and syntactic interactions peaked (avg. 53.12\%), while semantic interactions saw a modest rise (avg. 11.82\%). With only accessing the VACP interface in \replace{\reflmprinciple{S3}}{\reflmprinciple{S4}}, agents operated almost entirely at the semantic level (99.8\%).
Scenario \replace{\reflmprinciple{S3}}{\reflmprinciple{S6}} presented the most revealing patterns. With full access to all layers, agents favored semantic interactions (avg. 64.86\%) while still using a notable share of pragmatic interactions (avg. 29.04\%). They typically used semantic interactions for initial insights, then dropped to the pragmatic layer to visually verify solutions before answering.  Conversely, failed runs by GPT-5.2 in this scenario showed a collapse of this strategy with far fewer semantic interactions (25\%) and over-reliance on the syntactic layer (55\%).

\vspace{.25em}
\iviaparagraph{Qualitative Strategies and Sensemaking Quirks} Beyond aggregate statistics, the qualitative traces reveal notable sensemaking behaviors. \cref{fig:exampleWorkflow} shows a successful \replace{\reflmprinciple{S4}}{\reflmprinciple{S6}} run by Claude Sonnet 4.5, demonstrating a robust semantic-to-pragmatic loop: the agent selects the required year, queries the data semantically, and validates intermediate steps with screenshots.
In visually constrained environments (\reflmprinciple{S1}), agents used creative hacks to bridge the Gulf of Execution. When Claude Sonnet 4.5 failed to perform reliable brush selection, it inferred the subset from SVG geometries and tried to visually confirm its selection by mutating DOM styles to recolor lines. Gemini 3 Pro Preview (\reflmprinciple{S1}) estimated weather patterns by falling back to pixel counting via color thresholds. With semantic access (\replace{\reflmprinciple{S3}}{\reflmprinciple{S4}}), agents exhibited high-level analytical reasoning: Claude Sonnet noted the risk of visual misinterpretation and computed regression slopes directly from the data. But semantic access also introduced new pitfalls, such as GPT-5.2 failing by assuming it should only find IQR-based outliers and missing a second data point. In mixed environments, agents showed human-like corroboration, using SQL to identify the top country and then triggering a semantic hover tooltip to visually reconcile a rounding discrepancy.

\vspace{.25em}
\iviaparagraph{Design Implications for Agentic VA} These behaviors signal a key shift in how we must design interfaces for AI operators. Agents' tendency to add pixel-approximation layers or alter the DOM for visual confirmation suggests that VA systems should expose intermediate analytical states instead of forcing agents to infer them from visuals. The effectiveness of the semantic-to-pragmatic verification loop indicates that purely ``headless'' VA is inadequate: agents need to confirm semantic queries with visual representations, as humans do when comparing tables with charts. Future agent-ready VA systems should therefore offer a dual-channel architecture: a robust, API-driven semantic layer for reliable execution, paired with deterministic visual feedback to support agents’ internal evaluation and sensemaking.

\vspace{.25em}
\iviaparagraph{Summary} Our results show that a tailored semantic interface like VACP is not merely an enhancement but a requirement for reliable autonomous VA, bridging the gap between pixel-oriented web agents and the data-centric needs of VA applications.

\section{Discussion}
\remove{VACP introduces a semantic stratum that bridges the gap between complex data environments and agentic reasoning. 
By shifting from model-specific prompt engineering to structured context engineering, }VACP allows AI agents to interact with VA interfaces using the same mixed-initiative paradigm as human users \add{\cite{monadjemi_agentBased_2023}} \add{while maintaining human insights and control~\cite{dhanoa_agenticVis_2025} by shifting from model-specific prompt engineering to structured context engineering.}
Our evaluations show that the structured exposure significantly increases task completion. 
As frontier models converge in reasoning and tool-calling, the critical bottleneck for agentic VA becomes how well systems expose semantic state and affordances\remove{, instead of model quality}. 
Process analyses reveal that agents can recover from local coding errors but fail predictably when lacking interface context. 
VACP addresses this by encoding system knowledge and interaction options into an agent-tailored format, providing a robust gateway for accurate analytical execution. 
\remove{By offering a tailored, agent-agnostic toolset, }VACP significantly reduces token usage relative to baseline approaches, lowering costs for future systems. 
\remove{The underlying principles of VACP are highly generalizable (see \mbox{\cref{sec:VACP}}), enabling agent integration across broader data-rich environments, such as geographic information systems and 3D information modeling.} 
Preliminary experiments with smaller, on-device LLMs using VACP also showed promising success, opening the door for local AI agents\remove{to collaborate with humans in the same interface}.

%
%
%
%
%
%

%
%
%

%
%
%
%

\vspace{.25em}
\iviaparagraph{Limitations and Future Work}
While \replace{VACP establishes}{establishing} a robust foundation for agent-accessible VA, \replace{its}{VACP's} current design operates under several inherent conceptual and technical limits. 
\remove{Restricting exposure of the visual interface to semantic and structural context discards rich visual nuances.
 Spatial clustering, Gestalt principles, and dense visual patterns are easily perceived by humans but often lost when converted to text. 
Our experiments indicate that giving agents visual context in addition to the semantic VACP state 
, such as interface screenshots, improves performance, suggesting that multimodal access to layer
is crucial when token consumption is not a constraint.} 
\remove{Implementing VACP adds developer overhead. 
Complex, dynamically reconfigurable VA systems require careful mapping of application state and interactions to the protocol. 
For highly custom or rapidly changing interfaces, defining and maintaining these mappings can become a bottleneck.}
\remove{Future work will rigorously evaluate VACP on non-standard, highly dynamic VA systems to identify boundary conditions and scalability limits. 
Although VACP minimizes token usage and optimizes memory, its real-time synchronous collaboration can be bottlenecked by LLM inference latency, 
motivating research on asynchronous agent operations and interface designs that gracefully handle processing delays.}

\noindent\add{%
\textit{\textbf{Proof-of-Concept Evaluation:}} VACP-enabled agents show high success rates, but our benchmark relies on elementary analytical tasks to isolate core perception and execution failures. Agents that struggle here are not ready for real-world, more complex VA scenarios. We therefore do not claim that VACP currently supports complex exploratory workflows; it instead provides the operational foundation. Future work will rigorously evaluate VACP on non-standard, highly dynamic VA systems to determine its boundary conditions and scalability under greater visual and interactive complexity.%
}  
\noindent\add{\textit{\textbf{Visual Nuance:}} 
Visual nuances like Spatial clustering, Gestalt principles, and dense visual patterns are often lost, when restricting exposure to semantic and structural context. 
Our experiments indicate that giving agents visual context in addition to the semantic VACP state (\reflayer{L3} and \reflayer{L4}), such as interface screenshots, improves run-time and task completion performance, suggesting that multimodal access to layer~\reflayer{L1} is crucial when token consumption is not a constraint.
}  
\noindent\add{\textit{\textbf{Adoption Effort and Generalization:}} 
Implementing VACP adds developer overhead. While our current TypeScript library provides adapters for common declarative frameworks, complex, dynamically reconfigurable and custom VA systems require mapping of application state and interactions to the protocol. Further, the VACP library is designed for web-based systems. Applying VACP beyond web environments will require development of new transport layers and adapters. 
We envision the core principles of VACP generalizing to data-rich environments like GIS or 3D modeling, as they face, like VA, dynamic app states, available interactions, and precise execution~\cite{jankowski_3Denv_2013}.
}


\vspace{.25em}
\iviaparagraph{Research Opportunities}
Exposing VA interfaces through a structured, semantic protocol opens several research directions:

\noindent\add{\textit{\textbf{Autonomous Agentic Analysis:}} VACP can enhance autonomous agentic analysis beyond pure data access by providing access to available interactive VA systems. In future, it is relevant to explore how agents adapt their analysis processes given the additional modality.}
\noindent\textit{\textbf{Multi-Agent and Mixed-Initiative Systems:}} VACP supports concurrent access to a single application state, enabling mixed-initiative environments where specialized agents (e.g., data-cleaning, or analyst agents) collaborate seamlessly with human users. Future work can examine how this shared state affects user behavior, clarifying the benefits and drawbacks of such collaboration.
\noindent\remove{\textbf{Intelligent Agent-Driven Co-Adaptive Guidance:} VACP provides the infrastructure to advance co-adaptive guidance in VA. By enabling intelligent agents to continuously perceive interface state and reason about user interactions, it supports a shift from rigid, rule-based guidance systems (e.g., Lotse~\mbox{\cite{sperrle2023Lotse}}) to flexible, context-aware assistance tailored to the user's analytical flow.}%
\noindent\add{\textit{\textbf{Agent-Driven Co-Adaptive Guidance:}} By enabling intelligent agents to continuously perceive interface state and reason about user interactions, VACP supports a shift from rigid, rule-based guidance systems (e.g., Lotse~\cite{sperrle2023Lotse}) to flexible, context-aware assistance tailored to the user’s analytical flow.
}%
\noindent\textit{\textbf{Automated Testing of Analytical Provenance:}} Beyond standard functional UI testing, the protocol’s structured exposure of system states lets AI agents systematically navigate VA tools, enabling automated testing of complex data mappings and assessing the usability of analytical provenance tracking in highly interactive environments\add{, expanding works such as TraVIS~\cite{filosa_TraVis_2025}}.%
\noindent\remove{\textbf{Benchmarking Interactive VA:} To evaluate \add{the proof-of-concept implementation of} VACP, we curated a \remove{representative }set of interactive VA tasks. Moving forward, the community would benefit from a standardized benchmark, analogous to VLAT, specifically for assessing AI agent reasoning, interaction performance, and tool use in interactive VA interfaces.
}
\noindent\add{\textit{\textbf{Benchmarking Interactive VA:}} Moving forward, the community would benefit from a standardized benchmark, analogous to VLAT, specifically for assessing AI agent reasoning, interaction performance, and tool use in interactive VA interfaces.
}
\section{Conclusion}

We presented VACP, the first context protocol to provide AI agents native access to interactive VA applications\replace{. By}{, by} formalizing minimal requirements for both the interactive VA environment and the agent\remove{, VACP creates a principled bridge between complex visual interfaces and agentic workflows. We instantiated this framework as an open-source TypeScript library, \textit{\texttt{VACP}}}.
\\
Across diverse agent setups and \remove{representative}VA tasks, \add{our proof-of-concept} VACP \add{implementation} significantly outperforms state-of-the-art web-based AI agent interaction methods in reliability and performance. \replace{Beyond overcoming current technical limitations}{Further}, VACP serves as a \remove{core }building block for future mixed-initiative systems, enabling AI agents to seamlessly navigate, interpret, and manipulate VA interfaces alongside human users. To foster its use within the community, \replace{VACP}{our implementation} is available as open source software\textsuperscript{\ref{foot:github-link}}.

\acknowledgments{%
This work was partially funded by Julius Bär and foryouandyourcustomers. The compute credits for the experiments have been provided by Google and Microsoft. %
}

\bibliographystyle{abbrv-doi-hyperref}

\bibliography{references}

\clearpage
\appendix 
\section*{Analysis of Residual Errors}
\label{appendix:residualerrors}
Description of Errors:
\begin{itemize}
 \item \textbf{Incomplete Visual Data Exploration:} The agent fails to visually identify/find the required data point(s) in the visualization.
 \item \textbf{Visual Data Extraction:} The agent tries to read the required value(s) from the rendered UI (visualization, tooltips, axis, legends, etc.) but fails to retrieve the correct value from the pixels.
 \item \textbf{Failed Data Reverse-Engineering:} The agent tries to retrieve the required value(s) by parsing the DOM or reverse engineering, e.g., svg-elements, but fails to retrieve the correct value(s).
 \item \textbf{Interaction Errors:} The agent fails by interacting either with the wrong components or the interactions do not lead to the correct results.
 \item \textbf{Data Inspection:} The agent fails to request the right data when using the \textit{inspect\_data()} functionality.
 \item \textbf{Data Misinterpretation:} The agent misinterprets perceived or retrieved data values and provides an incorrect answer.
 \item \textbf{Non-Parsable Response:} The agent provides a response that is not parsable by the validator.
 \item \textbf{Validation Edge-Case:} The agent in general provided the correct answer, but the response did not satisfy the semantic validator.
 \item \textbf{Timeout:} The agent terminates without any response.\\
\end{itemize}

\begin{table}[h!]
  \vspace{-1.75em}
  \caption{%
    \add{Distribution of Error Types of the Failed Agent Runs per Ablation }%
  }
  \vspace{-1em}
  \label{tab:vis_papers}
  \scriptsize%
  \centering%
  \renewcommand{\arraystretch}{1} 
  \begin{tabularx}{\linewidth}{>{\linespread{0.75}\selectfont}X*{6}{w{c}{0.5cm}}}
    \toprule
    & \multicolumn{6}{c}{\textbf{scenario/ablation}}\\
    \add{\textbf{error type}} & \reflmprinciple{S1} & \reflmprinciple{S2}   & \reflmprinciple{S3}   & \reflmprinciple{S4} & \reflmprinciple{S5} & \reflmprinciple{S6} \\
    \midrule
    \add{Incomplete Visual\newline Data Exploration } & \add{24\%} & \add{28\%}  & \add{0\%} & \add{0\%} & \add{27\%} & \add{0\%} \\
    \add{Visual Data \newline Extraction} & \add{23\%} &     \add{23\%}  & \add{0\%} & \add{0\%} & \add{40\%} & \add{0\%} \\
    \add{Failed Data\newline Reverse-Engineering} & \add{9\%} & \add{6\%}  & \add{0\%} & \add{0\%} & \add{13\%} & \add{20\%} \\
    \add{Interaction Errors} & \add{13\%} & \add{13\%}  & \add{13\%} & \add{0\%} & \add{9\%} & \add{0\%} \\
    \add{Data Inspection} & \add{12\%} & \add{16\%}  & \add{44\%} & \add{50\%} & \add{0\%} & \add{40\%} \\
    \add{Data Misinterpretation} & \add{1\%} & \add{1\%}  & \add{6\%} & \add{50\%} & \add{0\%} & \add{20\%} \\
    \add{Non-Parsable Response} & \add{8\%} & \add{3\%}  & \add{0\%} & \add{0\%} & \add{9\%} & \add{20\%} \\
    \add{Validation Edge-Case} & \add{1\%} & \add{1\%}  & \add{13\%} & \add{0\%} & \add{0\%} & \add{0\%} \\
    \add{Timeout} & \add{9\%} & \add{9\%}  & \add{24\%} & \add{0\%} & \add{2\%} & \add{0\%} \\
    \midrule                
    \add{\makecell[l]{\linespread{0.75}\selectfont Total Errors/Runs}} & \add{90/180} & \add{88/180}  & \add{16/135} & \add{6/180} & \add{45/135} & \add{5/180} \\
    \bottomrule
  \end{tabularx}%
\end{table}

\end{document}